\documentclass[reprint,aps,prb,english,nolongbibliography,superscriptaddress]{revtex4-2}
\usepackage[colorlinks=true, urlcolor=blue, linkcolor=blue, citecolor=blue, pdftex]{hyperref}
\usepackage{xr} 

\usepackage[dvipsnames]{xcolor}
\usepackage[utf8]{inputenc}
\usepackage[normalem]{ulem}
\usepackage{braket}
\usepackage{graphicx}
\usepackage{caption}
\usepackage{caption}
\usepackage{subcaption}
\usepackage{multirow}

\usepackage{amsmath}
\usepackage{amsfonts} 
\usepackage[capitalise]{cleveref}

\newcommand{\bs}[1]{\boldsymbol{#1}}

\begin{document}

\title{Semiclassical picture of the Heisenberg spin glass in two dimensions: from weak localization to hydrodynamics}

\date{\today}

\begin{abstract}
The two-dimensional Heisenberg spin-glass model is investigated by means of a semiclassical expansion around classical states. At leading order, we obtain an effective quadratic spin-wave Hamiltonian and study the localization properties of its spectrum and eigenfunctions. We find that the nature of the spin-wave excitations, whether they are hydrodynamic or localized modes, depends crucially on the relevance/irrelevance -- in the renormalization group sense -- of the correlations induced by the underlying classical order in the spin-wave Hamiltonian matrix elements: low-energy excitations around magnetically ordered states are delocalized, whereas those around spin-glass ordered states are localized, albeit weakly. We interpret this phenomenology by relating the spontaneous breaking of spin-rotation symmetry in the original Heisenberg model to the symmetry and universality class of the resulting quadratic spin-wave Hamiltonian. We conjecture that the hydrodynamic picture can be recovered through the inclusion of interactions among the spin-wave excitations at higher order in the semiclassical expansion, favoring the onset of ergodic behavior.
\end{abstract}

\author{Giacomo Bracci Testasecca}
\email{gbraccit@sissa.it}
\affiliation{SISSA, via Bonomea 265, 34136, Trieste, Italy}
\affiliation{INFN, Sezione di Trieste, Via Valerio 2, 34127 Trieste, Italy}

\author{Jacopo Niedda}
\email{jniedda@ictp.it}
\affiliation{The Abdus Salam ICTP, Strada Costiera 11, 34151 Trieste, Italy}

\author{\\ Aldo Coraggio}
\affiliation{SISSA, via Bonomea 265, 34136, Trieste, Italy}
\affiliation{INFN, Sezione di Trieste, Via Valerio 2, 34127 Trieste, Italy}

\author{Roderich Moessner}
\affiliation{Max Planck Institute for the Physics of Complex Systems, Nöthnitzer Str. 38, 01187 Dresden, Germany}

\author{Antonello Scardicchio}
\affiliation{The Abdus Salam ICTP, Strada Costiera 11, 34151 Trieste, Italy}
\affiliation{INFN, Sezione di Trieste, Via Valerio 2, 34127 Trieste, Italy}

\maketitle

\section{Introduction}

Understanding how ergodicity emerges in presence of disorder in quantum systems represents a central and longstanding problem in modern condensed matter physics. Considering Heisenberg spin glasses (SG), since the seminal work of Halperin and Saslow \cite{Halperin1977}, it is believed that  hydrodynamics describes the low energy excitations, with Goldstone spin waves arising from the spontaneous breaking of the full $SO(3)$ symmetry group, in close analogy with the Heisenberg ferromagnet (FM) and antiferromagnet (AFM) \cite{Halperin_1969}. In particular, the hydrodynamic theory predicts three polarizations for the spin waves, related to the the fact all three generators of the rotational symmetry group are broken in a disordered spin-glass state. But hydrodynamics is rooted in the validity of ergodic dynamics close to an equilibrium state and, in particular in low dimensions, disorder is known to disrupt ergodicity in quantum systems through the mechanisms of Anderson Localization (AL) and Many-Body Localization (MBL) \cite{Anderson_1958,basko2006metal,EversMirlin_2008,Sierant_2025}. 

Localization phenomena are particularly striking in one and two dimensions, so the question arise: how can one assume the validity of hydrodynamics in presence of disorder in such systems? If, in some approximation, AL or MBL is at work, what is the way out? What are the timescales, if ever, on which hydrodynamics is valid?

To answer these questions, we further develop the semiclassical theory of the two-dimensional Heisenberg spin-glass model, which was recently addressed in~\cite{viteritti2025}, but dates back to~\cite{Bray_1981}. Remarkably, and similarly to what occurs in the case of the antiferromagnet~\cite{Anderson_1952,Oguchi_1960,Castilla1991}, the leading order term in the $1/S$ expansion predicts the values of the order parameter within a few percent accuracy for $S=1/2$. Besides validating the numerical evidence of a spin-glass phase in the ground state, this suggests that non-interacting spin-wave theory contains most of the information about the ground state of the model. The implications of this picture for dynamics and transport were beyond the scope of~\cite{viteritti2025} and we address them in this paper.

Although the symmetry class of the quadratic spin-wave problem is easy to identify, the correlations in the Hamiltonian matrix elements prevent one to conclude anything \emph{a priori} on its universality class, in the sense of the Altland-Zirnbauer (AZ) classification scheme \cite{AltlandZirnbauer1997}. These correlations arise from the minimum energy condition on the classical state around which the semiclassical expansion is performed and from the \emph{bosonic} Bogoljubov-de Gennes structure of the Hamiltonian \cite{Gurarie2003}. The main outcome of our analysis is that in the SG phase the whole spin-wave spectrum is weakly localized, as would happen for an uncorrelated random matrix theory with the same symmetries in $d=2$, thereby hindering the definition of a wavenumber. We interpret this result in terms of renormalization group (RG) \emph{irrelevance} of the correlations in the spin-wave Hamiltonian in the SG phase, and we contrast it to the case in which the underlying classical state is FM or AFM ordered. In the latter cases, we find a mobility edge in the spin-wave spectrum separating delocalized low-energy excitations from localized ones at higher energy, and therefore one cannot neglect the correlations induced by the underlying, polarized, classical state. 

The spin-wave localization in the SG phase has some important conceptual consequences. First, it means that the large part of the correction to the Edwards-Anderson order parameter found in~\cite{viteritti2025} is not due to long-range correlations among linear spin-wave modes, which vanish in the thermodynamic limit, but rather to local rearrangements of the spin configuration from the classical state, due to quantum fluctuations. Second it means that, low energy modes at leading order in $1/S$, in two dimensions, where AL prevails, cannot be the basis of the delocalized hydrodynamic modes of Halperin and Saslow. One is led to conjecture then that to recover the Halperin and Saslow hydrodynamic prediction in the thermodynamic limit, it is necessary to include interactions among the spin waves (naturally occurring at higher orders in the semiclassical expansion). The only mechanism which could prevent the onset of ergodicity in presence of interactions is MBL (even though this is thought not to exist in $d=2$ \cite{Potter2015Universal,DeRoeck2017Stability, Thiery2018Many,Morningstar2022Avalanches}) but we will see that the ``upside-down" structure of the spectrum, with the localization length decreasing as the energy of the excitation increases prevents even this possibility to take place. 

The paper is organized as follows. In Sec.~\ref{sec:model}, we introduce the model and briefly describe the magnetic properties of the classical minima. In Sec.~\ref{sec:SWexp} we derive the quadratic spin-wave Hamiltonian and discuss its symmetries; we detail the Bogoljubov diagonalization procedure, with special attention devoted to the presence of zero modes and the choice of the quasiparticle vacuum state. The main results on the localization properties of the spin-waves are presented in Sec.~\ref{sec:localiz} and discussed in the renormalization group and finite size scaling language. The role played by spin-wave interactions in recovering the hydrodynamics picture is discussed qualitatively in Sec.~\ref{sec:interactions}. Finally, Sec.~\ref{sec:conclusions} contains a discussion on the results of this work and on ongoing and future directions of research.

\section{The model}
\label{sec:model}
The quantum Heisenberg spin-glass model is defined by the following Hamiltonian:
\begin{equation}
\label{eq:ham_model}
    \hat{H} = \frac{1}{2} \sum_{ij} J_{ij} \hat{\mathbf{S}}_i \cdot \hat{\mathbf{S}}_j \ ,
\end{equation}
where $\hat{\mathbf{S}}_i=(\hat{{S}}^x_{i},\hat{{S}}^y_{i},\hat{{S}}^z_{i})$ is the $S=1/2$ spin operator on site $i=1,\dots,N=L^2$. The exchange couplings $J_{ij}$ are random binary numbers distributed according to 
\begin{equation}
\label{eq:prob_couplings}
    P(J_{ij}) = (1-p) \delta(J_{ij} + 1) + p \delta (J_{ij} - 1) \ ,
\end{equation}
where $p \in [0,1]$ is the fraction of antiferromagnetic bonds. We work on the square lattice, with nearest neighbors interactions and periodic boundary conditions. 

The model is frustrated and $SU(2)$ symmetric and is therefore affected by the sign problem, which restricts Quantum Monte Carlo simulations to regimes with only small fractions of disordered bonds~\cite{sandvik1994}. Exact-diagonalization studies have suggested the possible presence of a quantum SG phase~\cite{oitmaa2001,arrachea2001}, but they are necessarily limited to very small system sizes. The first robust evidence of SG order in the quantum ground state of this model was obtained in Ref.~\cite{viteritti2025}, using a novel neural-network quantum-state approach that gives access to unprecedented system sizes. In that work, we also performed a semiclassical spin-wave analysis, following the literature on spin-wave theory of disordered media~\cite{Sherrington_1977,Walker1977,Takayama_1978,Ching1980,Bray_1981,Ching_1981,Barnes_1981,Canisius1981}, which supported the neural-network results. In the present paper we extend this semiclassical framework to the dynamical properties of the model. Let us also mention two related facts. First, the one-dimensional version of the model, which was already studied in the past, see \emph{e.g.}~Ref.~\cite{Stinchcombe_1988}, has recently attracted renewed interest in Refs.~\cite{fava2024,sibei2025}, although the existence of a quantum SG phase in one dimension remains debated. Second, at the mean-field level, the fully connected model has been known to possess a quantum SG phase since the seminal work of Bray and Moore~\cite{Bray_1980}; only recently has this result been established rigorously for $S=1/2$ in Ref.~\cite{Kavokine2024}. As usual, what of the mean-field scenario applies to finite-dimensional systems is an open question.

To set up the semiclassical expansion, one must first identify the relevant classical configurations around which quantum fluctuations have to be computed. Since the system has continuous spin-rotation symmetry, any kind of long-range order, whether FM, AFM, and SG, can occur only at $T=0$, by the Mermin–Wagner theorem \cite{Fernandez_1977}. The problem then reduces to finding the classical equilibrium configurations, \emph{i.e.}, the solutions of the equations of motion that minimize the classical energy function. In particular, we seek the global minima. Once these configurations are identified, one can systematically compute the $1/S$ corrections. In the next section, we address this first step, which, somewhat surprisingly, has received little attention in the literature.

\subsection{Classical minima}

\begin{figure*}[t]
\centering
\begin{subfigure}{.49\textwidth}
  \centering
\includegraphics[width=0.98\linewidth]{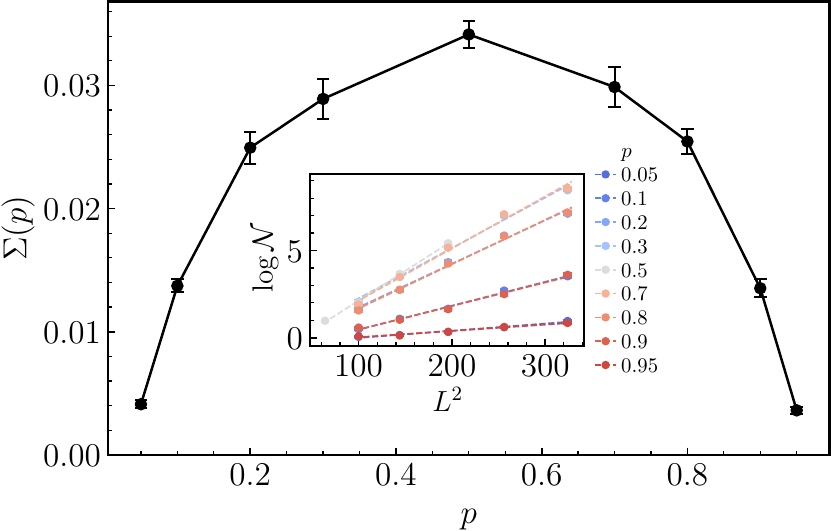}
\end{subfigure}
\begin{subfigure}{.49\textwidth}
  \centering
\includegraphics[width=0.98\textwidth]{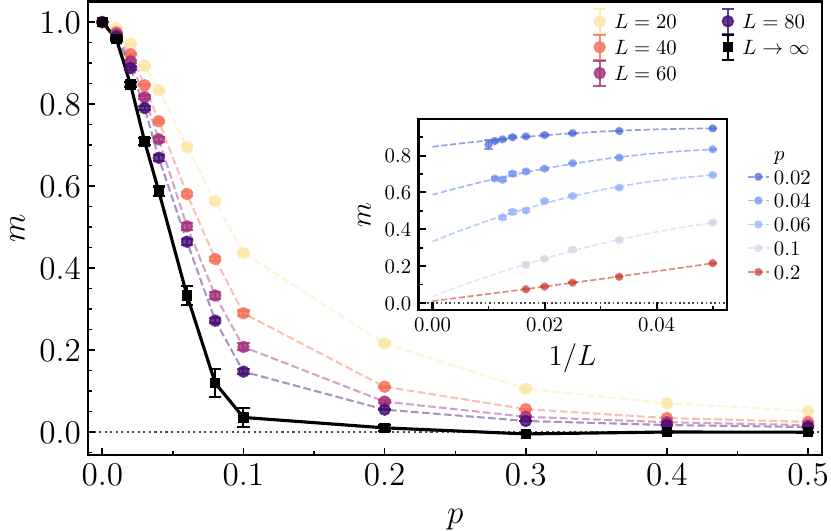}
\end{subfigure}
\caption{\textbf{Left panel}: Quenched energy complexity $\Sigma(p)$ as a function of the antiferromagnetic bond concentration $p$. $\Sigma(p)$ is computed as the linear fit slope of the disorder-averaged logarithm of the total number of minima of the Hamiltonian $\mathcal{N}$ as a function of $N=L^2$, shown in the inset. \textbf{Right panel}: Magnetization magnitude $m$ as function of $p$ for different system sizes (colored points and dashed lines) and large $L$ extrapolations (black points and continuous line). For each value of $p$, the extrapolated value of $m$ is obtained with a second order fit in $1/L$, as shown in the inset.}
\label{fig:mags}
\end{figure*}

The classical energy function is given by Eq.~\eqref{eq:ham_model}, where the spin operators are replaced by three-dimensional vectors $\mathbf{S}_i$ of unit norm, $|\mathbf{S}_i|^2 = 1$. In order to find the classical minima, we adopt a mixed microcanonical-canonical Montecarlo algorithm -- known as \emph{over-relaxation} -- borrowed from \cite{Baity2015soft,Baity2015inherent}, which combines gradient descent moves with energy conserving reflections, thereby enhancing a faster exploration of the energy landscape. The stationarity condition is given by $\mathbf{S}_i \propto \mathbf{h}_i $, where $\mathbf{h}_i = - \sum_{j \in \partial i} J_{ij} \mathbf{S}_j \; $ are the local fields. The algorithm stops when the total misalignment between the spin and the local field configurations reaches a fixed threshold. As a final check, we verify that the Hessian matrix computed around the final configuration is positive semidefinite, as expected for a local minimum.

Since the algorithm tends to become trapped in local minima, it is restarted several times from random initial configurations, to which we will refer to as \emph{replicas}. Different replicas may end up in the same minimum depending on the size of its basin of attraction. We observe that the probability of finding a minimum is typically larger for low-energy minima. The global minimum is then found more frequently than any other state, so finding it is difficult only because of {\it entropic} reasons, since there is an exponential number of equilibrium states.

In the left panel of Fig.~\ref{fig:mags}, the quenched energy complexity $\Sigma(p)$ is shown as a function of the antiferromagnetic bond concentration. $\Sigma(p)$ is defined as the logarithm of the \emph{total} -- \emph{i.e.}, summed over all energy values -- number of minima $\mathcal{N}(p)$ averaged over disorder realizations. To measure it, we limited this analysis to small system sizes ($L \leq 18$), where a sufficiently large number of replicas allows us to reliably sample all minima. We find that $\Sigma(p)$ is nonzero for all studied value of $p \in (0,1)$. A more detailed analysis of the classical energy landscape is beyond the scope of this work and will be presented elsewhere \cite{Coraggio2026}. Let us emphasize that even at small $p$, where the complexity is comparatively low, finding the global minimum remains challenging: for example, at $p=0.05$, we find $\Sigma(p=0.05) \simeq 0.004$, corresponding to order $10^6$ minima already at $L=60$. The situation becomes substantially worse at intermediate values of $p$, where the number of minima increases rapidly. Therefore, at the largest system sizes, no certificate of global optimality is available.

The most relevant property of the classical minima for the present analysis is whether they carry a magnetic signal. To characterize it, we measure the magnetization $\mathbf{m} = \tfrac{1}{N}\sum_i \mathbf{S}_i$ and the Néel order parameter $\mathbf{m}_{N} = \tfrac{1}{N} \sum_i (-1)^i \mathbf{S}_i$, averaged over disorder realizations. At the classical level, the value of the magnetization at $p$ is expected to coincide with that of the Néel order parameter at $1-p$ \footnote{One can obtain an AFM minimum of the classical energy from a FM one, by performing a staggered rotation, \emph{i.e.}~flipping all the spins on a sublattice, and changing sing to all the couplings, meaning that a minimum configuration at $p$ corresponds to another one at $1-p$. This symmetry is broken as soon as one takes $\hbar \neq 0$.}. Having checked that this symmetry is satisfied by our minimum configurations, we only show the magnetization magnitude in the right panel of Fig.~\ref{fig:mags}. The magnetization was already studied in Ref.~\cite{Cieplak_1985}, for system sizes up to $L=20$ and a few disorder realizations. We find that the magnetization extrapolates to zero at $p=0.2$. We also observe that the magnetization is approximately correlated with the energy of the minimum. Consequently, if the optimization selects higher-energy minima at large system sizes, the true ground-state magnetization may be systematically underestimated. 

At this point it is worth stressing that the robustness of the {\it classical} FM or AFM order against the addition of bond disorder in $d=2$ is in fact a subtle issue, whose solution is not among the scopes of the present work. In fact, while it would be natural to assume that a $p_c>0$ exists such that for $p<p_c$ the FM order is present, according to a comment by Villain in Ref.~\cite{Villain1979} (not brought up to a full argument in that paper and never mentioned again in the literature), a small density of defects can destabilize FM long-range order in two dimensions through the formation of dipolar distortions of the local spin orientation. However, even if this were the case, the associated magnetic domains size would diverge as $\ell\sim e^{c/p^2}$ and for $c=O(1)$ they would quickly become larger than the largest samples we considered between $p=0.1$ and $p=0.2$. Since we observe that for small enough values of $p$, the magnetization has an extremely slow dependence even when changing the system size by one order of magnitude (see the inset in the right panel of the Fig.~\ref{fig:mags}), we believe that for all practical purposes the FM and AFM phases we observe can be considered as proper thermodynamic phases.

SG order, instead, is a proper classical, thermodynamic state, confined at $T=0$. The Edwards-Anderson (EA) order parameter, defined as $Q^2 = \tfrac{1}{N^2} \sum_{ij} \left(\mathbf{S}_i \cdot \mathbf{S}_j \right)^2$, see \emph{e.g.}~\cite{Baity2015inherent,Binder_Young_1986}, extrapolates to $1/3$ in the thermodynamic limit for any value of $p$ for which the magnetic signal is absent~\footnote{In fact,
\begin{equation*}
    {\frac{1}{N^2}\sum_{i,j}\sum_{\alpha,\beta}S^\alpha_i S^\alpha_jS^\beta_i S^\beta_j=S^4\sum_{\alpha,\beta} \frac{1}{3}\delta_{\alpha\beta}\frac{1}{3}\delta_{\alpha\beta}=\frac{S^4}{3}}
\end{equation*}}. We denote by $p_1$ and $p_2$ the two values of the antiferromagnetic bond concentration at which for the explored sizes the magnetic and, respectively, the Neél order parameter extrapolate to zero, while the EA parameter remains finite. From the inset of the right panel of Fig.~\ref{fig:mags}, we have $p_1 \simeq 0.2$ and consequently $p_2 \simeq 0.8$. On the methodological side, deep in the SG phase, where the complexity is comparatively larger, which local minimum is selected to be the basis of the semiclassical expansion becomes less relevant, because the magnetization signal vanishes in all of them. This was already observed in Ref.~\cite{viteritti2025}.

\section{Spin-wave expansion}
\label{sec:SWexp}
To perform the semiclassical expansion we set $\hbar = 1$ and work in generic spin representation $S$, in order to take the limit $S \rightarrow \infty$. Spin-$S$ operators can be mapped to bosonic creation and annihilation operators by means of a standard Holstein-Primakoff (HP) map \cite{blaizot_ripka_1986}: $\hat{S}^z_i = S - \hat{a}^\dagger_i \hat{a_i},~ \hat{S_i}^+ = \sqrt{2S - \hat{a_i}^\dagger\hat{a_i}} \; \hat{a_i},~ \hat{S_i}^- = \hat{a_i}^\dagger \sqrt{2S-\hat{a}^\dagger_i \hat{a_i}} $,
where $\hat{a_i},\hat{a}^\dagger_i$ satisfy canonical commutation rules $[\hat{a}_i, \hat{a}_j^\dagger] = \delta_{ij}, \; [\hat{a}_i,\hat{a}_j] = [\hat{a}^\dagger_i,\hat{a}^\dagger_j] = 0$. The transformation preserves the spin algebra $[\hat{S}_i^\alpha,\hat{S}_j^\beta] = \delta_{ij}\epsilon_{\alpha \beta \gamma}\hat{S}^\gamma_i$, where Greek indices run over $\{x,y,z\}$ and $\hat{S}_i^\pm = \hat{S}^x_i \pm i \hat{S}^y_i$. At leading order in $1/S$, the HP map reads as
\begin{subequations}
    \begin{equation}
       \hat{S}^x_i \sim \sqrt{\frac{S}{2}}(\hat{a}_i + \hat{a}^\dagger_i) \; , \quad 
      \hat{S}^y_i \sim -i\sqrt{\frac{S}{2}}(\hat{a}_i - \hat{a}^\dagger_i) \; , 
    \end{equation}
    \begin{equation}
        \hat{S}^z_i \sim S - \hat{a}^\dagger_i \hat{a}_i \; .
    \end{equation}
\end{subequations}
Let us emphasize that, although the HP map is exact, the spin algebra is broken at each order in the large $S$ expansion.

Since the classical minima have randomly oriented spins, let us first perform a local change of reference with rotation matrices $R_i$ and rewrite the classical energy function as
\begin{equation}
\label{eq:class_ene}
    E_{cl} = \sum_{ij} J_{ij} R_i^{\beta \alpha } R_j^{t, \alpha \gamma} \bs{\Sigma}_i^\beta \bs{\Sigma}_j^\gamma \; ,
\end{equation}
where $R_i^{\alpha \beta} \bs{S}_i^\beta = \bs{\Sigma}_i^\alpha$, with $\bs{\Sigma}_i$ aligned along the $z$-axis for all sites, as in a \emph{clean} ferromagnet. In the previous equation, summation over repeated indices is assumed. By applying the HP map to the rotated spin operators $\hat{\bs{\Sigma}}_i$ and truncating the $1/S$ expansion at the leading order, one finds
\begin{equation}
    \frac{\hat{H}}{S^2} = E_{cl}  + \frac{1}{S} \hat H_{SW} + o(1/S)\ ,
    \label{eq:energyexp}
\end{equation}
where 
\begin{equation}
    \label{eq:HSW}
\hat H_{SW} = \frac{1}{2} \sum_{ij} \left(\hat{a}_i^\dagger A_{ij} \hat{a}_j + \hat{a}_i^\dagger B_{ij} \hat{a}_j^\dagger \right) + \text{h.c.} \; .
\end{equation}
Since the Hamiltonian $\hat{H}$ is of order $O(S^2)$, we divided it by a factor $S^2$ in Eq.~\eqref{eq:energyexp}, in order for $\hat{H}_{SW}$ to appear at order $O(1/S)$. The explicit expression of the matrix elements is given in Appendix~\ref{app:matrix_elements}. 

In order to have more compact expressions, we introduce the following notation \cite{blaizot_ripka_1986}. We define $2N\times 2N$ block matrices:
\begin{equation}
    \gamma = \begin{pmatrix}
0 & 1_N  \\
1_N & 0  \\
\end{pmatrix}~~~\eta = \begin{pmatrix}
1_N & 0  \\
0 & -1_N  \\
\end{pmatrix} \; ,
\label{eq:gammaetamatr}
\end{equation}
acting on $\hat{\bs{\alpha}} = (\hat a_1, \cdots,\hat a_N, \hat a^\dagger_1, \cdots, \hat a_N^\dagger)^t$. The matrix $\gamma$ implements charge conjugation: $\hat{\bs{\alpha}}^\dagger =  \gamma
\hat{\bs{\alpha}} $, while $\eta$ is a $2N$-dimensional metric, in terms of which the bosonic commutation relations can be written as $[\hat{\bs{\alpha}},\hat{\bs{\alpha}}^\dagger]=\eta$. It holds: $\gamma^2 = \eta^2 = 1_{2N}$. With this notation:
\begin{equation}
\label{eq:Ham_SW}
    \hat H_{SW} = \frac{1}{2} \hat{\bs{\alpha}}^\dagger M \hat{\bs{\alpha}}  -\frac{1}{2} \mathrm{t
    r} A \ ,
\end{equation}
where $M$ is a block matrix defined as:
\begin{equation}
    M = \begin{pmatrix}
A & B  \\
B^* & A^* \\
\end{pmatrix} \; .
\end{equation}

The spin-wave Hamiltonian $\hat H_{SW}$ describes a system of coupled harmonic oscillators in terms of the bosonic creation and annihilation operators $\hat a_i^\dagger, \hat a_i$. The number operators $\hat{n}_i = \hat{a}^\dagger_i \hat{a}_i$ measure the local deviations from the classical background and the corresponding amplitude of the fluctuations in the planes orthogonal to the classical spin orientations. One could also express the Hamiltonian in terms of position and momentum operators: $\hat x_i \equiv \hat S^x_i/\sqrt{S}$ and $\hat p_i \equiv \hat S^y_i/\sqrt{S}$, which at order $1/S$ satisfy the commutation relations $[\hat x_i, \hat p_j] = i \delta_{ij}$. In these variables the Hamiltonian has a coupling term between  position and momentum \cite{Gurarie2003}. We will return to this representation in the following sections, whenever it provides a more insightful physical description.

\subsection{Symmetry class}
\begin{table}[t]
\centering
\begin{tabular}{lcc |ccc}
\hline\hline
 & \multicolumn{2}{c|}{\textbf{Classical state}}
 &  & \multicolumn{2}{c}{\textbf{SW Hamiltonian}} \\
State & TRS & ~RRS
& $p$ & TRS & Charge \\
\hline
FM  & No  & $SO(2)$ & $0$       & Yes & $U(1)$ \\
AFM & No  & $SO(2)$ & $1$       & Yes & None  \\
SG  & No  & None     & $(0,1)$   & No  & None  \\
\hline\hline
\end{tabular}
\caption{Symmetry properties of the classical state for ferromagnetic (FM), antiferromagnetic (AFM), or spin-glass (SG) order and of the spin wave (SW) Hamiltonian for the clean FM/AFM case ($p=0,1$) and in presence of disorder ($p \neq 0,1$). TRS denotes time-reversal symmetry; RRS stands for  residual rotational symmetry in the classical state; charge refers to the possible global $U(1)$ particle number conservation in the spin-wave theory.
}
\label{tab:symmetries}
\end{table}

Before discussing the symmetries of the spin-wave problem, let us first spend a few words on those of the classical states and discuss what one may expect from them regarding the low-energy excitations. In generic spatial dimensions, the symmetry breaking pattern of the Heisenberg model is $SO(3) \rightarrow SO(2)$ in the FM and AFM phases and $SO(3) \rightarrow O(1)$ in the SG phase. The Goldstone mechanism generates a number of gapless excitations that is related to the number of broken generators of the group. As is known, the AFM has two Goldstone modes with linear dispersion relation, while the FM has only one Goldstone mode with quadratic dispersion relation, in line with the correct Goldstone mode counting in non-relativistic theories \cite{NIELSEN1976,Watanabe_2011,Watanabe_2012}. From this argument, one would expect three linearly dispersing Goldstone bosons in the SG phase, coinciding with the hydrodynamic modes discussed by Halperin and Saslow~\cite{Halperin1977}; however, it has to be noticed that a disordered state violates one assumption of the Goldstone theorem, by \emph{completely} breaking translational invariance (TI), thereby impairing the very definition of a wavenumber. 

Besides spin rotational symmetry, the classical state also breaks time reversal symmetry (TRS) in all three cases, since TRS acts on a spin by changing its sign \cite{landlif1980}. We notice that the \emph{clean} antiferromagnet ($p=1$) is peculiar for this symmetry, since TRS is microscopically broken for the Néel state, but can be recovered in combination with lattice translations.

Let us now discuss the symmetries of the non-interacting bosonic Hamiltonian in Eq.~\eqref{eq:Ham_SW}. The matrix elements defined in Eqs.~\eqref{eq:matrixelementsOFF} satisfy $A_{ij}=A_{ji}^*$ and $B_{ij} = B_{ji}$, implying $\hat H_{SW}=\hat H_{SW}^{\dagger}$. Besides being hermitian, the Hamiltonian is invariant under charge conjugation 
\begin{equation}
    \gamma K \hat H_{SW} K^{-1} \gamma^{-1} = \hat H_{SW} \; ,
\end{equation}
where $K$ is the operator implementing complex conjugation. This symmetry is characteristic of Bogoljubov–de Gennes Hamiltonians, a notable example of which arise in the study of (disordered) superconductors with the pairing treated in the mean-field approximation \cite{Bogo_1947,Gurarie2003, Seibold2012, Seibold2015}, the key difference here being the \emph{bosonic} nature of the degrees of freedom. 

We notice that the structure of the matrix elements of $\hat{H}_{SW}$ is directly tied to the underlying classical spin configuration. In particular, when two neighboring spins are non-collinear in the classical state -- which happens for all values of $p \neq 0,1$ -- the corresponding matrix element acquires a complex phase. As a consequence, the Hamiltonian is not invariant under complex conjugation alone, \emph{i.e.}~$K \hat H_{SW} K^{-1} \neq \hat H_{SW}$, and TRS is explicitly broken. By contrast, in the \emph{clean} limits where the classical configuration is fully polarized (\emph{i.e.}~all spins are collinear), all matrix elements are real and TRS is preserved. From these considerations, we conclude that the model we are studying has the  symmetries of class D in the Altland-Zirnbauer (AZ) classification scheme \cite{AltlandZirnbauer1997,Xu2020}.

However, it has to be noticed that this does not mean that the spin-wave problem is in the same universality class, since AZ classes comprise \emph{random matrices with uncorrelated entries}, as the more traditional Wigner-Dyson ones \cite{Mehta2004}. The Hamiltonian in Eq.~\eqref{eq:Ham_SW} has instead correlated random elements, due to the fact that it has been obtained by expanding around a minimum. A first source of correlations is the positive-definiteness of $\hat H_{SW}$, as always is for generic bosonic systems \cite{Gurarie2003}. Most importantly, the matrix elements of $A$ and $B$ carry within their correlations the local structure of the classical minimum, which is encoded in the local rotation matrices $R_i$ appearing in Eq.~\eqref{eq:class_ene}.  

We conclude this section by reminding the reader that $B=0$ for the clean ferromagnet ($p=0$), leading to a $U(1)$ invariant $\hat{H}_{SW}$. The particle number conservation in the pure ferromagnet is what makes this case exceptional and is related to the fact that the order parameter is itself a constant of motion, \emph{i.e.}~$[\hat H, \hat{\mathbf{S}}^z]=0$, where $\hat{\mathbf{S}}^z$ is the $z$-component of the total spin. As a consequence, the classical order is protected by this symmetry and receives no corrections from quantum fluctuations. In terms of the position and momentum variables $\hat x_i$ and $\hat p_i$ introduced before, $B=0$ means that there is no direct coupling between $\hat x_i$ and $\hat p_i$, and the Hamiltonian can be diagonalized by a simple Fourier transformation. 

The discussion about the symmetries of the classical state and of the spin-wave Hamiltonian is summarized in Table~\ref{tab:symmetries}.

\subsection{Bogoljubov diagonalization}

The quadratic problem defined in Eq.~\eqref{eq:Ham_SW} can be solved with a Bogoljubov transformation \cite{Colpa_1978,blaizot_ripka_1986,Bogo_1947}. We define \emph{quasiparticle} operators $\hat{\bs{\beta}} = (\hat b_1, \cdots,\hat b_N, \hat b^\dagger_1, \cdots, \hat b_N^\dagger)^t$, which are connected to the original bosonic operators by a linear transformation: $\hat{\bs{\beta}} = T \hat{\bs{\alpha}}$. As usual, we require $T$ to be symplectic, \emph{i.e.}~$T\eta\gamma T^t = \eta \gamma$, so that $[\hat{\bs{\beta}},\hat{\bs{\beta}}^\dagger]=\eta$, and unitary, \emph{i.e.}~$T = \gamma T^* \gamma$, leading to the following condition
\begin{equation}
    T \eta T^\dagger \eta = 1_{2N} \ .
\end{equation}
This equation can be used to write an explicit expression of the matrix $T$ and of its inverse:
\begin{equation}
T = \begin{pmatrix}
X^* & -Y^*  \\
-Y & X  \\
\end{pmatrix} \qquad
 T^{-1} = \begin{pmatrix}
X^t & Y^\dagger \\
Y^t
& X^\dagger  \\
\end{pmatrix} \ ,
\label{eq:BogandInv}
\end{equation}
from which we can derive a form of the canonicity condition that is the straightforward generalization of the TI case, in the original theory of superconductors \cite{Bogo_1947}:
\begin{equation}
    \begin{split}
    XX^\dagger - YY^\dagger = 1_N  \qquad X^\dagger X - Y^t Y^* = 1_N \\
    X Y^t -Y X^t = 0_N \qquad X^t Y^*  -Y^\dagger X=0_N    \ .     
    \end{split}
\label{eq:comm_rel}
\end{equation}

Then, by expressing $\hat H_{SW}$ in terms of the quasiparticle operators, we realize that $T$ is not exactly the matrix diagonalizing $M$, but rather $\eta M$:
\begin{equation}
\label{eq:Bogo-Ham generic}
    \hat H_{SW} = \frac{1}{2} \hat{\bs{\beta}}^\dagger \eta \Omega \hat{\bs{\beta}} - \frac{1}{2} \text{tr} A \ ,
\end{equation}
with 
\begin{equation}
    \Omega = T D T^{-1}, \qquad D=\eta M \ .
\end{equation}
The matrix $D$ is called \emph{dynamical matrix} \cite{Colpa_1978}, it is non-Hermitian in general and its presence represents the main difference with respect to \emph{fermionic} Bogoljubov-de Gennes systems. The diagonalization procedure is affected by mathematical subtleties regarding the kernel of $D$, which are discussed in Appendix~\ref{app:bogo}.

\section{Localization of the spin waves}
\label{sec:localiz}

In this section we  study the localization properties of the spin waves. This is the central result of this paper, which can summarized as follows: despite the symmetry class being the same (D in the AZ classification scheme), the correlations of the underlying classical state make the localization properties completely different for the three different phases: FM, SG, and AFM.

\begin{figure}
    \centering
\includegraphics[width=0.98\linewidth]{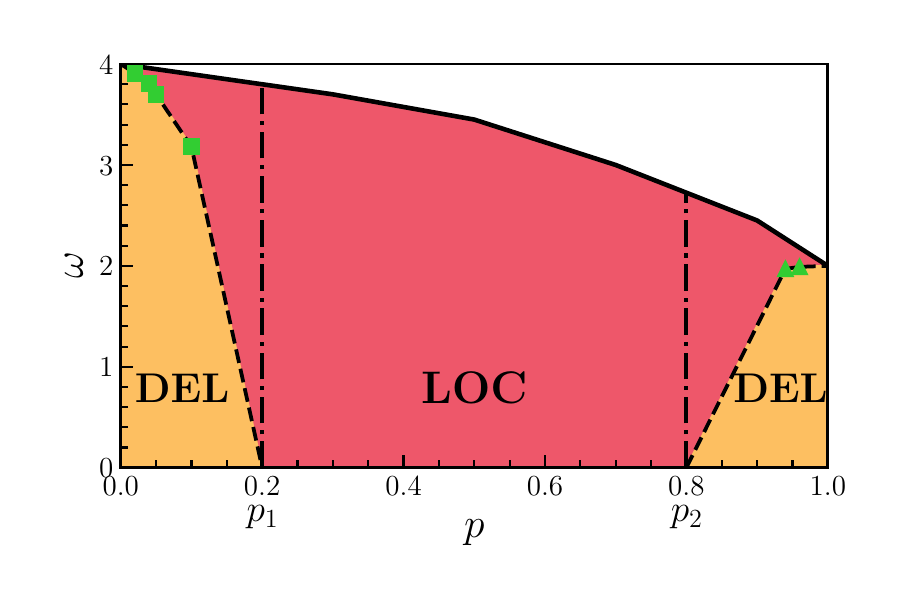}
    \caption{Phase diagram of the spin-wave Hamiltonian. The continuous lines mark ground state transitions between the FM/AFM phases to the SG phase in the classical solution. The dashed lines mark a mobility edge between localized and extended regions inside the ordered phases. The points correspond to the mobility edge $\omega_c(p)$, as computed in Appendix~\ref{app:mobedge}.}
    \label{fig:phasediag}
\end{figure}

For a classical SG state the whole excitation spectrum is localized, albeit \emph{weakly}, according to the naive expectation in $d=2$. For classical FM and AFM states, for $p < p_1 $ or $p > p_2$, the signal of the magnetization and of the Néel order parameter respectively create a delocalized region at low energy. There is a \emph{mobility edge} $\omega_c(p)$, such that modes with energy $\omega<\omega_c(p)$ are delocalized, while modes with energy with energy $\omega>\omega_c(p)$ are localized. The dynamical phase diagram of the model is sketched in Fig.~\ref{fig:phasediag}. It is important to stress that this is the opposite of what happens in the Anderson model, where the localized states are at {\it lower} energies. This is a manifestation of Goldstone's theorem and is important when one introduces  interactions among the spin waves. We will give a qualitative description of the effect of the interactions in Sec.~\ref{sec:interactions}. The presence of a mobility edge in the low-energy excitation spectrum of a disordered model was already discussed in Ref.~\cite{Canisius1981}.

\begin{figure}
    \centering
    \includegraphics[width=0.98\linewidth]{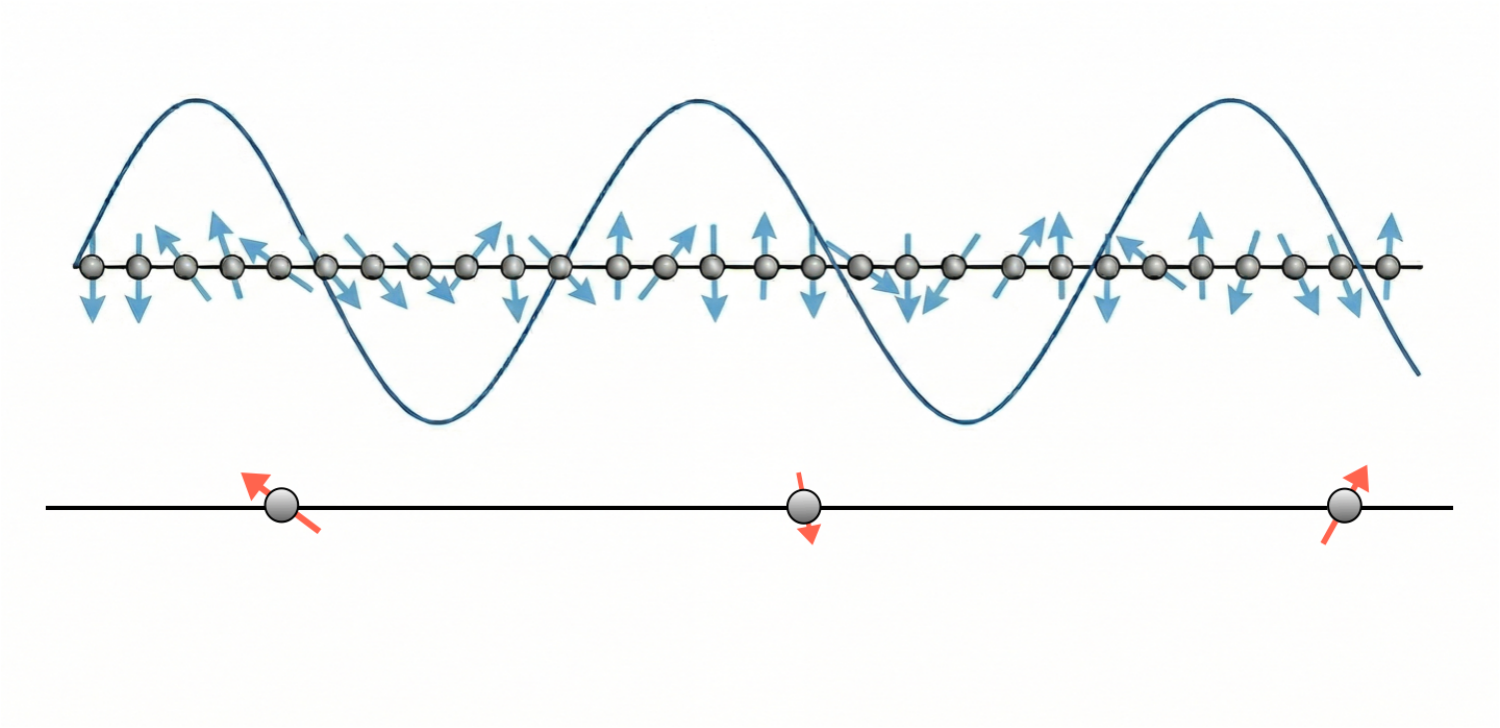}
    \caption{Coarse graining the underlying classical spin system to long wavelengths $\lambda$ reduces the local random fields by a factor of $1/\sqrt{\lambda}$ for every spatial dimension. In $d=2$ the reduction is $1/\lambda$.}
    \label{fig:coarse_grain}
\end{figure}

There is a simple, heuristic argument in support of this observation regarding why higher energy modes ``feel" a stronger disorder. At any fixed $p$, for any wavelength $\lambda$ of the spin wave, we can think of coarse-graining the system to lengths of $O(\lambda)$, below which the wave cannot resolve details  (see Fig.~\ref{fig:coarse_grain}). In this way, the effective fields acting on the effective spins in a box $B(\mathbf{r},\lambda)$ of size $\lambda$, around the point $\mathbf{r}$ are random vectors $\mathbf{h}(\mathbf{r},\lambda)\simeq\sum_{i\in B(\mathbf{r},\lambda)}\mathbf{h}_i$. Remembering that $\mathbf{h}_i\propto \mathbf{S}_i$ of the classical solution, in the SG phase these contributions are random, with average zero and amplitudes $h(\lambda)\sim \Delta h\sqrt{\lambda^{2}}$, where $\Delta h^2$ is the variance of the microscopic disorder strength, and $\Delta h^2 \propto J^2$, \emph{i.e.}~the variance of the random coupling distribution. However, the effective spin grows like $\lambda^2$, which is the number of spins in the block \footnote{The effective spin does not correspond to the coarse graining of a (disordered) spin configuration, but of the spins as \emph{dynamical} variables, canonically conjugated to the rotation field. As such, each spin in a box of size $\lambda$ gives to the effective field an $O(1)$ contribution.}. We see then that the effective disorder is $W\sim J \lambda^{-1}$. Since the dispersion relation of the spin waves in the SG phase is $\omega\propto J\lambda^{-1}$ (again following \cite{Halperin1977}), the disorder ``felt'' by a wave is exactly proportional to the frequency of the wave, $W \propto \omega$. Similar considerations made for FM and AFM states need to take into account the fact that the underlying classical spin configuration is correlated to give a signal $\sum_i\mathbf{h}_i \propto \mathbf{m}$ and $\sum_i(-1)^i\mathbf{h}_i \propto \mathbf{m}_N$, which are both proportional to $\lambda^2$ and not $\sqrt{\lambda^2}$. 

Summarizing, we expect the low-energy excitations of FM and AFM states to be as delocalized as their clean counterparts ($p=0,1$), while those of SG states to suffer from the presence of a disorder proportional to their frequency.

\subsection{Observables}

To characterize localization properties, we consider two standard observables, one based on spectral statistics and the other one on eigenstate properties, both normalized to take values in the interval $[0,1]$. 

As a spectral probe, we employ the so-called $r$-parameter \cite{Oganesyan2007}. Given an ordered set of eigenvalues $\omega_\alpha$, one first defines the level spacings $\delta_\alpha = \omega_{\alpha+1} - \omega_\alpha$, and then constructs the ratio between adjacent spacings as
\begin{equation}
   r_\alpha = \frac{\min(\delta_\alpha,\delta_{\alpha+1})}{\max(\delta_\alpha,\delta_{\alpha+1})} \; . 
\end{equation}
Averaging over small energy windows $[\omega, \omega + \Delta \omega]$ and disorder realizations at a given size $L$ yields the mean value $r(\omega,L)$. This quantity takes universal values in the thermodynamic limit depending on the nature of the spectrum. For uncorrelated (Poisson) statistics, characteristic of localized systems, one finds $r_{\mathrm{P}} = 2\ln 2 - 1 \simeq 0.386$. By contrast, for chaotic systems described by random matrix theory in the Gaussian Unitary Ensemble (GUE), one obtains $r_{\mathrm{GUE}} \simeq 0.5996$ \cite{Atas2013}, which also corresponds to the ergodic fixed point value for systems in the AZ class D. For convenience, we introduce the rescaled quantity
\begin{equation}
    \phi = \frac{r - r_{\mathrm{P}}}{r_{\mathrm{GUE}} - r_{\mathrm{P}}} \; ,
    \label{eq:resc_r}
\end{equation}
which interpolates between $\phi=0$ in the localized (Poisson) limit and $\phi=1$ in the fully ergodic (GUE) regime.

As an eigenstate-based diagnostic, we consider the fractal dimension $D$ \cite{EversMirlin_2008}, defined as 
\begin{equation}
    D = \frac{\partial S}{\partial \ln L^2}  \; ,
    \label{eq:fractD}
\end{equation}
where $S$ denotes the participation entropy, which will be defined shortly. For non-interacting systems, this quantity captures how broadly a given eigenmode is distributed in space (\emph{i.e.}~over the lattice), and is therefore sensitive to localization properties. In the present setting, however, its definition requires additional care due to the Bogoljubov structure of the eigenmodes. In fact, one must keep in mind that Bogoljubov modes are described by two-component vectors living in a space with double the sites of the lattice, corresponding to non-commuting degrees of freedom (\emph{e.g.}, $\hat x_i$ and $\hat p_i$). Therefore, they can not be naively interpreted as standard single-particle wavefunctions, whose square yields a well-defined probability according to the Born rule \footnote{From this point of view, Bogoljubov modes resemble many-body wavefunctions living in the Fock space, and this reflects the absence of particle number conservation. In fact, the same difficulties in studying eigenstate localization in real space also are found in the many-body case, even if in that case they are more severe.}. As a consequence, defining a properly normalized spatial distribution -- commonly employed to characterize eigenstate localization in non-interacting systemsbecomes nontrivial. To circumvent this issue, we adopt a physically motivated definition based on the local increase of the oscillator \emph{action}. Specifically, we define the quantity
\begin{equation}
    R_\alpha(i)^2 = \mathcal{R}^{-1}_\alpha \left(\braket{\alpha| \hat a^\dagger_i \hat a_i |\alpha} - \braket{0| \hat a_i^\dagger \hat a_i |0}\right),
    \label{eq:prob_depl}
\end{equation}
which measures the increase in the action of the oscillator $\hat a_i$ (in units of $2\pi \hbar$), due to the creation of a quasiparticle mode $\ket{\alpha} = \hat b_\alpha^\dagger \ket{0}$ of energy $\omega_\alpha$. The normalization factor $\mathcal{R}_\alpha$ is chosen such that $\sum_i R_\alpha(i)^2=1$, ensuring that $R_\alpha(i)^2$ can be consistently interpreted as a spatial weight associated with the mode $\alpha$. In Appendix~\ref{app:probR} we derive two complementary expressions for this quantity: one in terms of the position and momentum variables, $\hat x_i$ and $\hat p_i$, which provides a natural interpretation of $R_\alpha(i)^2$ as an increase in the action; the other in terms of the eigenstate components $X$ and $Y$, defined in Eqs.~\eqref{eq:BogandInv}, which is used for the numerical evaluation of $R_\alpha(i)^2$.

We can now define the participation entropy of an excitation $\alpha$ as
\begin{equation}
    S_\alpha=-\sum_i^N R_\alpha(i)^2\ln R_\alpha(i)^2,
\label{eq:part_ent}
\end{equation}
and, after averaging over states in small energy windows $[\omega,\omega+\Delta \omega]$ and disorder realizations, we get the entropy $S(\omega,L)$ and the fractal dimension $D(\omega,L)$ accordingly. In the case of magnon excitations for clean FM and AFM states, this definition gives $D(\omega,L)=1$, therefore characterizing a fully ergodic excitation.

\subsection{Ferromagnetic and antiferromagnetic phases: delocalization at low energies}

\begin{figure*}[t]
\centering
\begin{subfigure}{.47\textwidth}
  \centering
\includegraphics[width=0.98\linewidth]{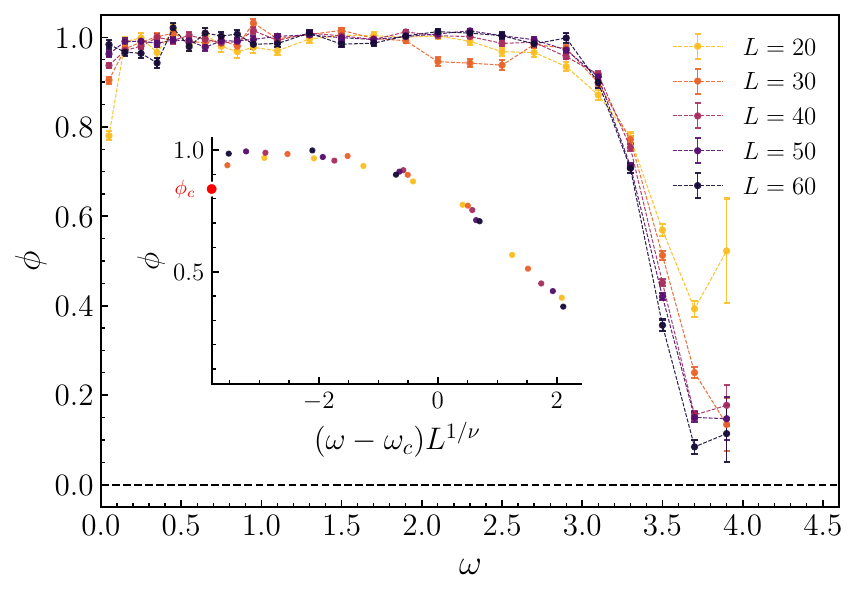}
\end{subfigure}
\begin{subfigure}{.49\textwidth}
  \centering
\includegraphics[width=0.98\textwidth]{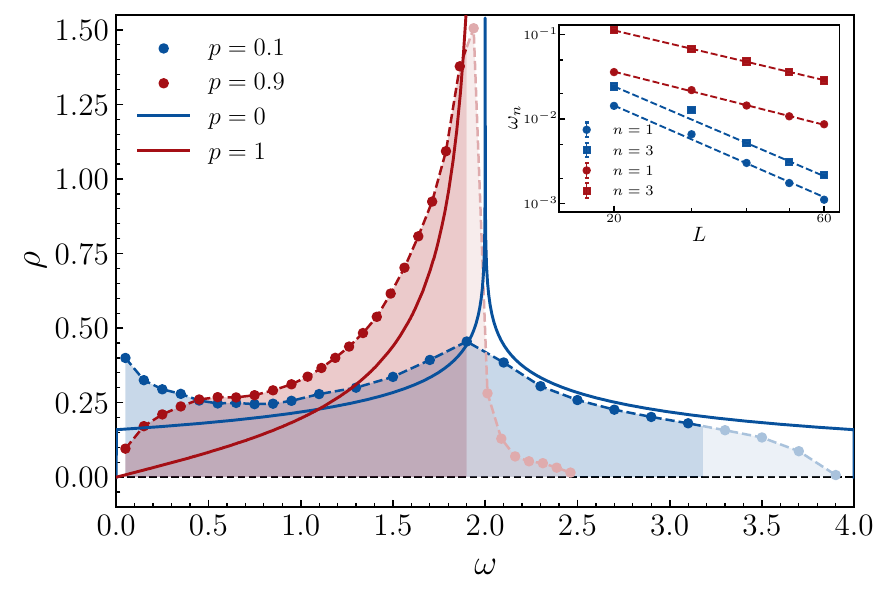}
\end{subfigure}
\caption{\textbf{Left panel}: Rescaled $r$-parameter $\phi$, Eq.~\eqref{eq:resc_r}, averaged over $\sim 1000$ disorder realizations at $p=0.1$, with FM signal. There is a critical energy $\omega_c$ separating low-energy delocalized from high-energy localized modes. The inset contains a data collapse obtained with $\nu = 2.1$ and $\omega_c = 3.18$, yielding a $\phi_c = 0.84$. \textbf{Right panel}: Density of states for two different values of antiferromagnetic bond concentration, $p=0.1$ and $p=0.9$ (points and dashed lines), compared to their clean counterparts, $p=0$ and $p=1$ (continuous lines, computed analytically). Data pertains to the largest simulated size, $L=60$, and is averaged over $\sim 1000$ disorder realizations. Different filling colors distinguish delocalized modes (darker) from  localized ones (lighter). The scaling of the averaged $n$-th lowest eigenvalue with system size is shown in the inset. The values of the dynamical critical exponent $z$ extracted respectively for $n=1$ and $n=3$ are: $z=2.27(5), 2.25(3)$ at $p=0.1$, and $z = 1.21(2),1.23(1)$ at $p=0.9$.}
\label{fig:DOS_FSS}
\end{figure*}

As discussed before, the spin excitations in the FM and AFM phases show a mobility edge. This is characterized by a critical energy $\omega_c(p)$ which vanishes as $p\to p_1$ or $p_2$. At first sight, this appears to be in contradiction with the AZ classification, which assigns the system to class D. Indeed, bosonic class D systems are always localized in $d=2$. This apparent discrepancy is resolved by noting that the coefficients of the quadratic expansion in Eq.~\eqref{eq:HSW} are not {\it iid} random variables. Indeed, these coefficients, among other things, encode the presence of a finite magnetization (or Néel magnetization) in the underlying classical state. The existence of a preferred direction of spontaneous magnetization implies that an effective $SO(2)$ rotational symmetry is preserved at long wavelengths, thereby qualitatively modifying the localization properties of the low-energy excitations.

In a localization transition the system becomes scale invariant at the critical energy $\omega_c$. The critical exponent $\nu$ controls the scaling, the localization length diverges as $\xi \sim (\omega -\omega_c)^{-\nu}$ and its value is universal and fixed only by symmetry and effective dimensionality, related to the value of $p$. For $p=0.1$, we numerically extract both the critical value $\omega_c$ and the exponent $\nu$. This analysis performed on the rescaled $r$-parameter $\phi$ is shown in the left panel of Fig.~\ref{fig:DOS_FSS}. We perform a data collapse to find the best values of $(\omega_c,\nu)$: computing the reduced $\chi^2$ for several fitting functions and defining the confidence interval as the fits whose $\chi^2$ is lower than $1.05 \chi^2_{min}$,  we find $\omega_c \in [3.14,3.24]$ and $\nu \in [1.8,2.5]$. Other values of $p$ are discussed in Appendix~\ref{app:mobedge}.

In recent studies \cite{Jiricek2026UniversalRelation,vanoni2023renormalization} a \emph{super-universal} scaling relation between the critical values of the $r$-parameter and the fractal dimension in Anderson models have been conjectured, for both GUE and GOE (Gaussian Orthogonal Ensamble) symmetry classes. We check if this relation also holds in the present case for $p=0.1$, where we find $r_c \simeq 0.57$ (corresponding to $\phi_c \simeq 0.84$  in Fig.~\ref{fig:DOS_FSS}) and $D_c \simeq 0.75$, matching with the prediction in figure 5 of \cite{Jiricek2026UniversalRelation}. The value of the critical point $D_c$ is estimated with the same procedure as for $\phi_c$. The details are given in Appendix~\ref{app:mobedge}. This result is remarkable, considering that our system belongs to a different symmetry class with respect to GOE and GUE, \emph{i.e.}~bosonic AZ class D, in which the definition of the fractal dimension is affected by the issues discussed above. 

In the right panel of Fig.~\ref{fig:DOS_FSS}, the densities of states in the FM state at $p=0.1$ and in the AFM state at $p=0.9$ are compared with their clean counterparts. Despite belonging to the same symmetry broken phase, data suggest that low-energy properties of the weakly disordered systems are different from those of the clean ones. In particular, the presence of disorder induces a decrease of the dispersion relation, leading to an abundance of low-energy modes, according to the relation $\rho(\omega) \propto \int d^2 k~\delta(\omega - k^z)$, where $z$ is the dynamical exponent \footnote{Notice: this is just a proxy of $z$, whose correct value can be extracted from the correlation functions as a function of time. This is beyond the scope of this work and is left for following research.}. Although the system is not TI, one can define a dispersion relation $\omega \sim k^z $, in terms of a quasi-momentum $k$, and relate this to the scaling of system size $\omega \sim L^{-z}$. In the inset of Fig.~\ref{fig:DOS_FSS}, we show the scaling of the low-energy modes and extract the dynamical critical exponent. While only two eigenvalues are displayed, we have verified that the entire low-energy sector scales with the same exponent: $z \simeq 2.26$ for $p=0.1$ and $z \simeq 1.23$ for $p=0.9$. Notably, these values exceed those of the clean systems, indicating sub-diffusive transport in the FM phase (clean FM, $z=2$) and sub-ballistic transport for the AFM (clean AFM, $z=1$). As we discuss in Sec.~\ref{sec:interactions}, interactions are expected to restore the standard hydrodynamic behavior.

\subsection{Spin glass phase: Renormalization group scaling}

\begin{figure*}
\centering
\begin{subfigure}{.49\textwidth}
  \centering
\includegraphics[width=0.98\linewidth]{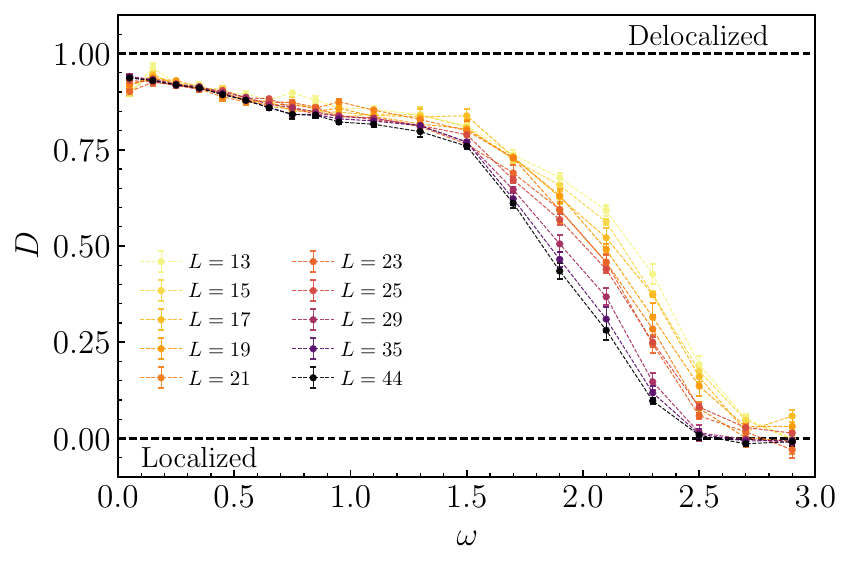}
\end{subfigure}
\begin{subfigure}{.49\textwidth}
  \centering
\includegraphics[width=0.98\textwidth]{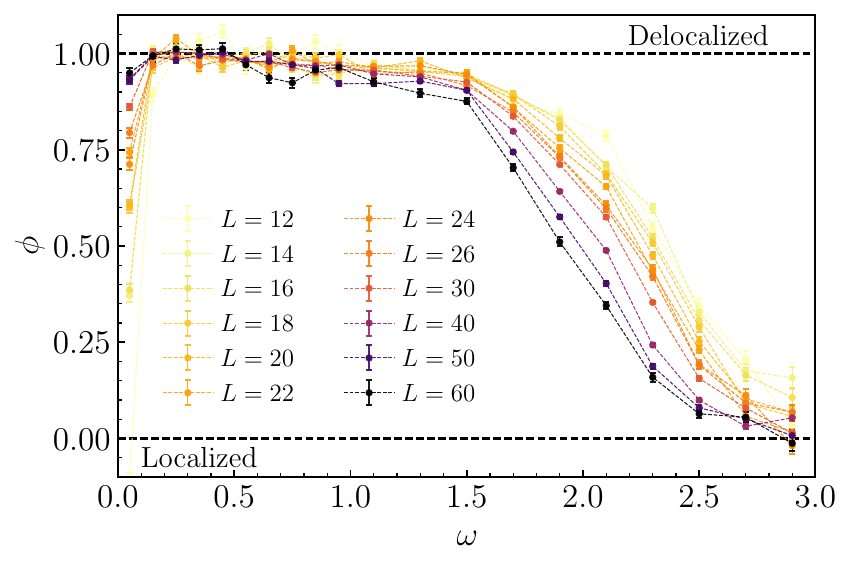}
\end{subfigure}
\caption{\textbf{Left panel}: Fractal dimension, defined in Eqs.~(\ref{eq:fractD}-\ref{eq:part_ent}), in the SG phase for $p=0.5$ averaged over $\sim 1000$ disorder realizations. The values of $L$ in the legend correspond to the geometric mean of the sizes in each block, where $D$ is computed through a linear fit of the participation entropy (see Appendix \ref{app:FractalDim} for more details). \textbf{Right panel}: Rescaled $r$-parameter $\phi$, Eq.~\eqref{eq:resc_r}, in the SG phase for $p=0.5$, averaged over $\sim 1000$ disorder realizations. In both panels data are shown as a function of the energy $\omega$ for various system sizes.}
\label{fig:DandR}
\end{figure*}

In the spin-glass phase, for $p \in (p_1,p_2)$, the spin-wave excitations are fully localized. However, this localization is \emph{weak}, in the sense that strong finite-size effects (logarithmic in system size) persist and simple diagnostics are not sufficient to capture the scaling behavior. For this reason, it is useful to adopt a more systematic framework based on scaling theory. 

A powerful approach to Anderson localization (AL) is provided by the \emph{one-parameter scaling} hypothesis \cite{Abrahams_1979}, which organizes the dependence of physical observables on the system size. This framework has been successfully applied in all finite spatial dimensions $d$ \cite{altshuler2025renormalization}, and only breaks down in the limit $d \to \infty$, which is the case of expander graphs \cite{vanoni2023renormalization}. More recently, the same ideas have been extended to many-body systems \cite{niedda2025,balducci2025}, where they helped in systematizing the finite-size effects which have puzzled researchers to the point of leading some of them to suggest MBL might actually not exist even in one dimensional models~\footnote{To cut a long story short, we consider the situation settled by the theorems of \cite{de2024absence} based on the work \cite{imbrie2016many}, but see also \cite{imbrie2017local,ros2015integrals} for the connections with the perturbation theory of \cite{basko2006metal}.}

The central assumption of one-parameter scaling is that, \emph{in the scaling limit} (\emph{i.e.}\ for sufficiently large system sizes), the behavior of the system is controlled by a single parameter. In the renormalization group (RG) language, this corresponds to the existence of a single relevant operator. One can take as such the dimensionless conductance $g$ \cite{Abrahams_1979} (with $g \rightarrow \infty$ in the ergodic phase and $g\rightarrow 0$ in the localized phase) or its inverse $t=1/g$ \cite{EversMirlin_2008} and write the beta function as
\begin{equation}
    \beta_t = - \frac{d t}{d \ln L^2} \; ,
\end{equation}
where $L$ is the largest scale in the problem (\emph{i.e.}~the system size). In the present case, which falls in the AZ symmetry class D, the conductance is the thermal one, since the particle number is not conserved, \emph{i.e.}~there is no electric charge. A key consequence of the one-parameter scaling hypothesis is that any observable in the scaling regime must be a smooth function of $t$, and, similarly, different observables must be analytically related to each other. We can then study the $\beta$ function of any of such quantities, and obtain any other by a change of variables (with the appropriate Jacobian). 

In absence of -- or on the irrelevance of -- the correlations in the matrix elements of the Bogoljubov-de Gennes Hamiltonian, we can expect that the phenomenology is that of symmetry class D. In $d=2$, the $\beta$ function for class D systems is always negative, reflecting the fact that these systems are localized even at infinitesimal disorder, and it approaches zero in the limit of vanishing inverse conductance, $t \to 0$. In particular, the prediction of the $2+\epsilon$ expansion in the non-linear sigma model describing class D systems is: $\beta_t \sim - t^{\alpha_t}$, with $\alpha_t=2$, for $\epsilon=0$ \cite{EversMirlin_2008}, similarly as for the Anderson model in $d=2$ with GOE symmetry class. 

While $t$ is the natural scaling variable to compare with field theoretical predictions, it is not directly accessible in our numerics. Instead, we measure the scaling variables $D$ and $\phi$, which are shown in Fig.~\ref{fig:DandR}, and whose beta functions are defined as
\begin{equation}
    \beta_D = \frac{d\ln D}{d \ln L^2} \; , \qquad  \beta_\phi = \frac{d\ln \phi}{d \ln L^2} \; .
    \label{eq:betafuncDphi}
\end{equation}
However, we do not know what is the relationship between the scaling variables $D,\phi$ and the inverse conductance $t$. For the Anderson model in $d=2$ it was argued in \cite{altshuler2025renormalization} (and amply supported numerically) that, close to the ergodic fixed point, as $t \rightarrow 0$ and $D \rightarrow 1$, then $1 - D \sim t$, thereby implying $\alpha_D = 2$. We will check whether the present case is compatible with this scenario in a moment.

\begin{figure*}
\centering
\begin{subfigure}{.49\textwidth}
  \centering
    \includegraphics[width=0.965\linewidth]{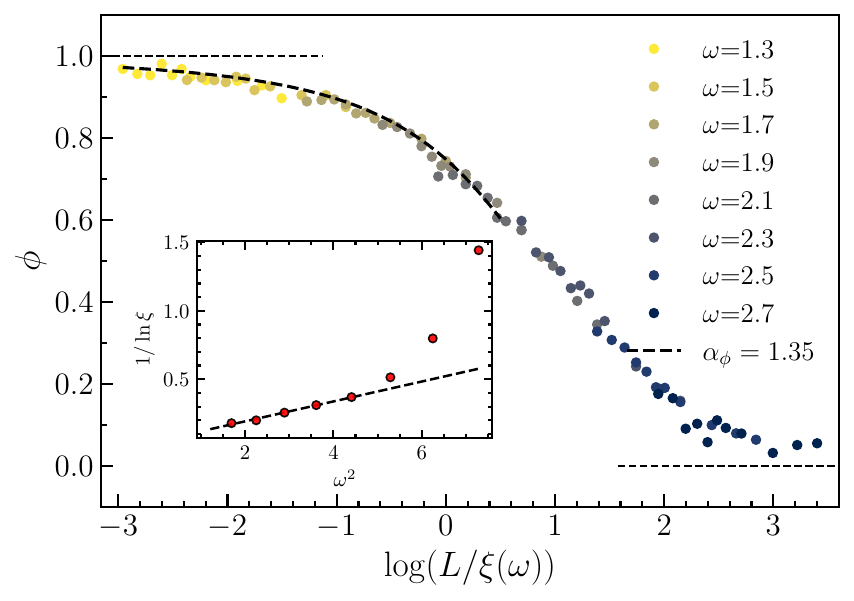}
\end{subfigure}
\begin{subfigure}{.49\textwidth}
  \centering
\includegraphics[width=0.98\textwidth]{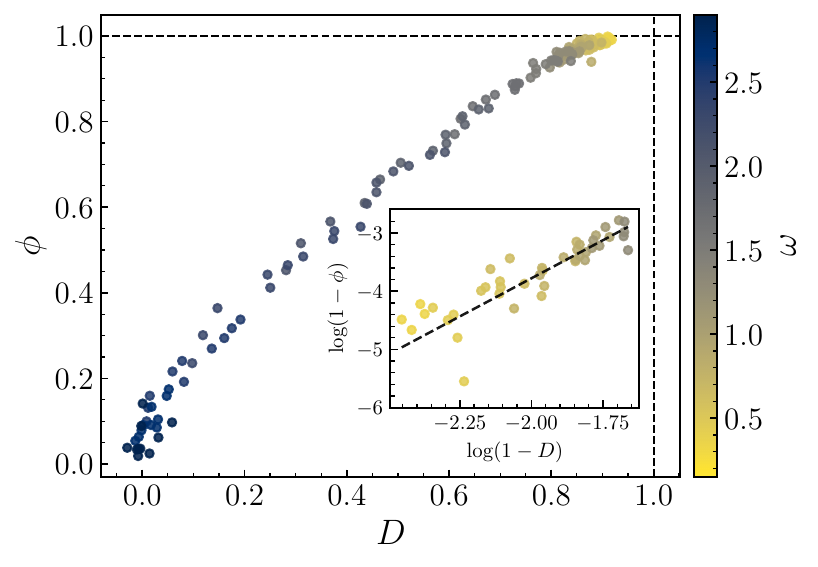}
\end{subfigure}
\caption{\textbf{Left panel}: Data collapse of the rescaled $r$-parameter $\phi$, Eq.~\eqref{eq:resc_r}. Each color corresponds to a different value of the energy $\omega$. The black dashed curve is obtained by inverting Eq.~\eqref{eq:integrated_beta} at leading order: $(1-\phi)^{1-\alpha_\phi} \sim \ln(L/\xi)$ with $\alpha_\phi = 1.35$, in order to find $\phi(\ln L/\xi)$. The inset shows the scaling of the localization length with energy, which follows the relation $\xi \sim e^{\omega_0^2/\omega^2}$, with $\omega_0 = 3.7(1)$ . \textbf{Right panel}: Rescaled $r$-parameter $\phi$ as a function of the fractal dimension $D$. The points aligning on a single curve is a validation of the one-parameter scaling hypothesis. Inset: log-log fit of $1-\phi$ vs $1-D$ near the ergodic fixed point. The slope of the fit yields a value of the exponent $\gamma$ defined in the text: $\gamma = 2.6(3)$, with an intercept $4.3(5)$.}
\label{fig:RGSCaling}
\end{figure*}

Let us first introduce a set of reasonable assumptions that will serve as an interpretative framework for our numerical results. A minimal hypothesis is that also $\beta_D$ and $\beta_\phi$ have power law behaviors near the ergodic fixed point:
\begin{subequations}
    \begin{align}
        \beta_D  &\sim - (1-D)^{\alpha_D} \; , \qquad D \sim 1 \; , \label{eq:betaD} \\
        \beta_\phi &\sim - (1-\phi)^{\alpha_\phi} \; , \qquad \phi \sim 1 \; , \label{eq:betaphi} 
    \end{align}
\end{subequations}
with some universal exponents $\alpha_D$ and $\alpha_\phi$. Moreover, we assume that in the same scaling regime the two observables $D,\phi$ are related as
\begin{equation}
    1-\phi \sim (1-D)^\gamma \; ,
\end{equation}
for some exponent $\gamma>1$, possibly non integer. In that case, the relationship between the $\beta$ functions of $D$ and $\phi$ is simple and one can easily derive an equation connecting the three exponents $\alpha_D, \alpha_\phi$ and $\gamma$:
\begin{equation}
\label{eq:exponents}
    \frac{\alpha_D}{\gamma} = \frac{1}{\gamma} - 1 + \alpha_\phi \; . 
\end{equation}
We plan to test this relations with the values of the exponent extracted from our numerics.

Let us now turn to the numerical analysis. First, we directly extract $\beta_D$ (see Appendix~\ref{app:FractalDim}) and find that $\alpha_D$ is compatible with the value $\alpha_D = 2$. Next, we estimate $\alpha_\phi$ by integrating the flow equation for $\phi$ near the ergodic regime, \emph{i.e.}~for $\phi \sim 1$, see Eq.~\eqref{eq:betaphi}:
\begin{align}
      &2\ln(L/L_0) = \int_{\phi(L_0)}^{\phi(L)} \frac{d \phi}{\phi \beta_\phi}\nonumber\\
      &\sim \left. \frac{(1-\phi)^{1-\alpha_\phi} }{1-\alpha_\phi} F\!\left(1,\,1-\alpha_\phi;\,2-\alpha_\phi;\,1-\phi\right)\right|_{\phi(L_0)}^{\phi(L)} \; ,
      \label{eq:integrated_beta}
\end{align}
and fitting the data as a function of $\ln( L/\xi)$ with the functional form above, where $F$ is the Gauss hypergeometric function and $\xi$ is defined in such a way as to absorb all the dependence on $L_0$ and $\phi(L_0)$. We obtain a good data fit for values of $\alpha_\phi$ in the interval $[1.2, 1.5]$. In Fig.~\ref{fig:RGSCaling} we plot the scaling of the variable $\phi$ and the theoretical curve obtained by inverting Eq.~\eqref{eq:integrated_beta} at leading order for $\alpha_\phi = 1.35$. For bosonic random systems in two dimensions it is expected that the localization length diverges as $\xi \sim e^{\omega_0^2/\omega^2}$, for $\omega \to 0$  \cite{Gurarie2003,John1983}. In the inset of Fig.~\ref{fig:RGSCaling} (left panel), we verify this scaling and extract the value of $\omega_0$, which turns out to be $\omega_0 = 3.7(1)$. Finally, we study the relation between $\phi$ and $D$, shown in the right panel of Fig.~\ref{fig:RGSCaling}, and extract the exponent $\gamma$ as the slope of the logarithmic fit of $1-\phi$ vs $1-D$ (shown in the inset), finding $\gamma = 2.6(3)$. By using this value of $\gamma$ and $\alpha_D=2$ to compute $\alpha_\phi$ through Eq.~\eqref{eq:exponents}, we obtain $\alpha_\phi \in [1.3,1.4]$, which is compatible with the interval estimated directly from the data for $\phi$. This provides a validation of the one-parameter scaling relation in Eq.~\eqref{eq:exponents} and confirms the belonging of the model in the SG phase to the AZ class D.

\subsection{Renormalization group scaling around the critical points}
\begin{figure}
    \centering
    \includegraphics[width=0.98\linewidth]{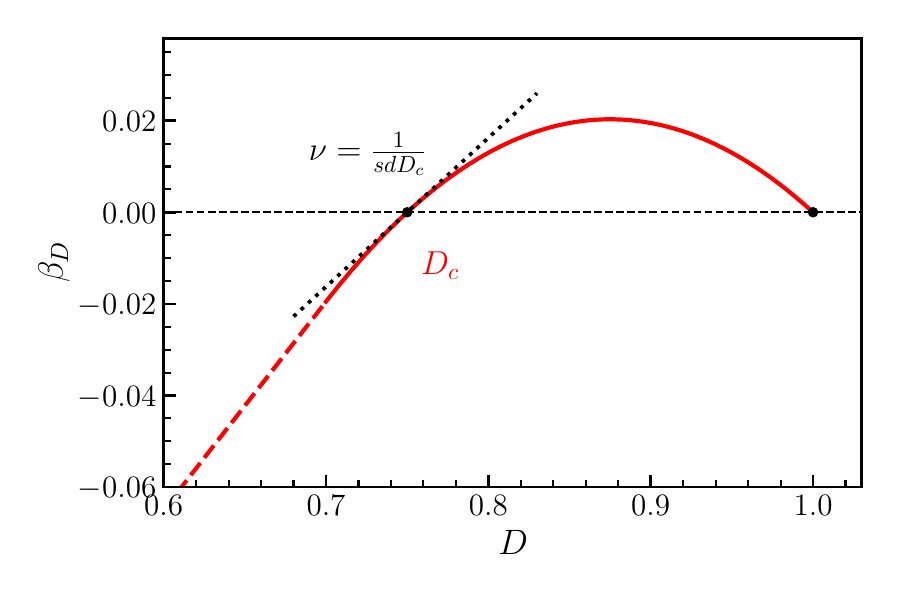}
    \caption{Plot of the one-parameter scaling $\beta$ function for the FM at $p=0.1$, see Eq.~\eqref{eq:betaFM}. The critical exponent $\nu$ is inversely proportional to the slope $s$ of the $\beta$ function at the critical point.}
    \label{fig:betafunction_cartoon}
\end{figure}

The RG analysis performed in the previous section can be extended to FM/AFM states, where, as discussed above, the correlations induced in the Hamiltonian matrix elements by the residual $SO(2)$ symmetry of the classical state lead to a delocalization transition in the lower part of the spectrum. In this case the beta function must have a zero in a finite value $D_c$ and an attractive ergodic fixed point. To capture this phenomenology, we make the simplest assumption on the beta function form for FM/AFM states, close to the critical points:
\begin{equation}
    \beta_D \simeq - c_1 (1-D)^2 + c_2 \frac{p_c - p}{p_c} (1-D) \; ,
    \label{eq:betaFM}
\end{equation}
where $p_c=p_1$ or $p_2$ and $c_1$ is the coefficient of the beta function in the SG phase, which we assume to be independent on $p \in (p_1,p_2)$. By using the values: $c_1 = 1.3(1)$ at $p=0.5$ (see Appendix~\ref{app:FractalDim}) and $p_c = p_1 \simeq 0.2$, we solve the equation $\beta_D = 0$ for $p=0.1$ and $D_c \simeq 0.7$ and extract the coefficient $c_2 = 0.60(5) $. From this, we can compute the critical exponent $\nu$ as \cite{altshuler2025renormalization}: $ \nu = 1/(s d D_c)$, where $d=2$ and $s=c_2 \frac{p_c-p}{p_c}$ is the slope of the beta function in $D=D_c$ (which coincide with the slope in $D=1$, for symmetry). We find a value of $\nu = 2.0(2) $, compatible with the one extracted by the data of $\phi$, see Fig.~\ref{fig:DOS_FSS}. This is an important consistency check of our numerical analysis. With these values of the coefficients $c_1$ and $c_2$, we plot the beta function for $p=0.1$ in Fig.~\ref{fig:betafunction_cartoon}.

\section{$1/S^{3/2}$ corrections and restoration of ergodicity}
\label{sec:interactions}

We have seen in the previous sections that the $1/S$ expansion to lowest order gives a mobility edge separating delocalized and localized excitations in the FM and AMF states, and a fully localized excitation spectrum in the SG phase. Upon the introduction of $J/S$ corrections, which act as interactions between the spin waves, the three kinds of order behave in different ways. 

While we do expect that the FM and AFM states will go quickly to ergodicity on local timescales $ \tau \sim S/J$, with possibly some patches of relatively slow dynamics coming from the high-energy localized modes, the situation is completely different in the SG phase. The whole spectrum is localized, albeit weakly, and the interaction should first overcome the localization of the noninteracting modes to restore ergodicity. This situation is parallel to that encountered in Basko, Aleiner, and Altshuler (BAA) \cite{basko2006metal} and in \cite{ros2015integrals} except for one fundamental difference: there can be no upper bound to the localization length $\xi(\omega)$ as function of the mode energy $\omega$. This is a fundamental property of the Goldstone modes, which changes the picture completely. In this section, which is the only speculative part of this paper, we shall picture the consequences of the interactions between the modes, which will be present in the full $S=1/2$ Heisenberg model and, consequently, in experiments.

The interaction terms, in units in which the quadratic Hamiltonian is $O(S^0)$, are of the form
\begin{equation}
    \hat H_{3}\sim\sum_{i,j}\frac{J}{\sqrt{S}} \hat a^\dagger_i \hat a_i \hat a^\dagger_j+\mathrm{h.c.} \; ,
\end{equation}
where $i,j$ are nearest neighbors on the lattice. After passing to the Bogoljubov modes $\hat b_\alpha$ this becomes
\begin{equation}
    \hat H_{3}\sim\frac{J}{\sqrt{S}}\sum_{\alpha,\beta,\gamma}M_{\alpha,\beta,\gamma} \hat b^\dagger_\alpha \hat b^\dagger_\beta \hat b_\gamma+\mathrm{h.c.} \; ,
\end{equation}
where the tensor element $M_{\alpha,\beta,\gamma}$ contains the information on the mode structure and can be written in terms of the Bogoljubov matrices $X,Y$.

If all the modes were localized by a localization length $\xi=O(1)$ (\emph{i.e.}\ if there is an upper bound to the localization length $\xi(\omega)$) then this processes could be treated perturbatively and, in the approximations of \cite{basko2006metal,ros2015integrals}, one could argue for an MBL phase (however all the caveats of dealing with $d=2$ apply \cite{Potter2015Universal,DeRoeck2017Stability, Thiery2018Many,Morningstar2022Avalanches}). If this were the case, only a subtle, non perturbative effect $O(e^{-S})$ would restore ergodicity and hydrodynamics. However, if there is no upper bound and $\xi\sim e^{\omega_0^2/\omega^2}$ for small $\omega$, this argument cannot be used. 

\begin{figure}
    \centering
    \includegraphics[width=0.9\linewidth]{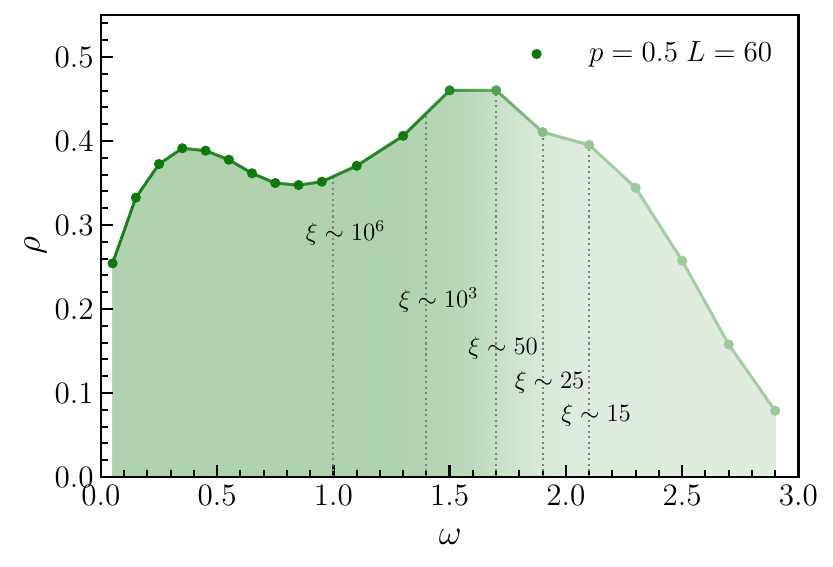}
    \caption{Density of states at $p=0.5, \; L=60$. The shading represents the finite size crossover between localized and delocalized states. A darker color marks a finite size extended state, where $\xi_{loc} \geq  60$ which is the largest system size explored in this paper. The dashed vertical lines are the finite size crossover points, identified by $\xi_{loc} = L$. }
    \label{fig:dos_P05}
\end{figure}

In the present case, the situation is much more similar to the following scenario. The single-particle spectrum is split in two by an energy $\omega_c$: the states with $\omega>\omega_c$ are localized with localization length $\xi$, and the states with $\omega<\omega_c$ are \emph{effectively} delocalized spin waves. For example one can take $\omega_c$ such that $\xi(\omega_c)=L$, the system size. This would give $\omega_c=\omega_0/\sqrt{\ln L}$, with $\omega_0\simeq 3.7$ from the numerics. We make a little error, if any, to consider this a simple quantity of $O(1)$, independent of $L$. For example for $L=100$, $\omega_c\simeq 1.7$ and for $L=10^6$, $\omega_c\simeq 1.0$. Using the data in Fig.~\ref{fig:dos_P05}, this corresponds to a fraction of localized excitations $P\simeq 2/3$ and delocalized excitations $1-P\simeq1/3$. In the following, we will refer to the delocalized excitations as spin waves, and to the localized ones as localized states. Moreover, for simplicity, we will assume that the delocalized modes are simple plane waves with some momentum $\mathbf{q}$ and energy $\omega(q)$.

\begin{figure}
    \centering
    \includegraphics[width=1\linewidth]{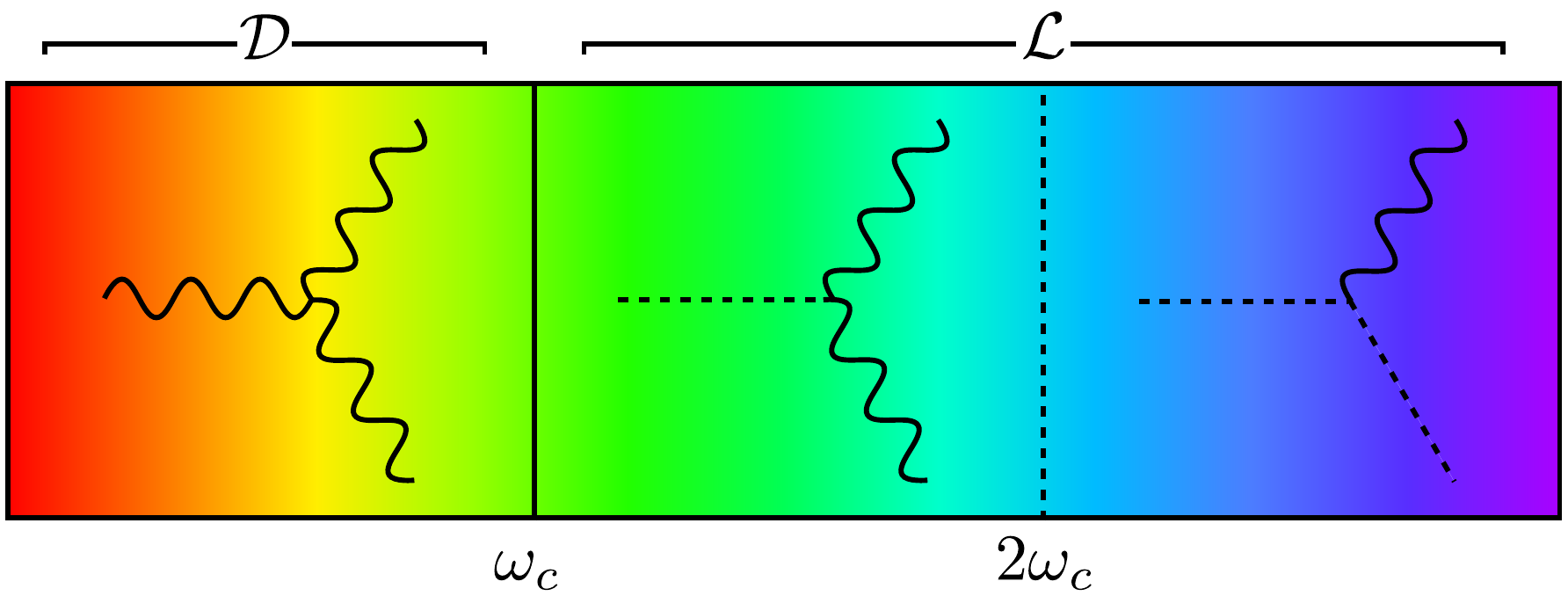}
    \caption{The open decay channels of excitations, depending on the energy of the initial state. From left to right (low energy to high energy), delocalized to two delocalized, localized to two delocalized, and localized to localized and delocalized modes. Dashed/wiggly lines represent localized/delocalized excitations.}
    \label{fig:decay_channels}
\end{figure}

By classifying the modes $\alpha\in\mathcal{L}$ or $\alpha\in\mathcal{D}$ (whether their localization length is smaller or larger than the system size $L$) we have now different possible cases which need to be treated (see Fig.~\ref{fig:decay_channels}). 

If the $\alpha,\beta,\gamma$ modes are $\mathcal{L}$, as we said above, we are back in a BAA/MBL scenario where ergodicity cannot be restored, or if it does, the reason is the dimensionality of space \cite{de2016absence,chandran2016many}. If the $\alpha,\beta,\gamma$ modes $\in\mathcal{D}$ then this is a typical matrix element of spin wave interactions and it will restore ergodicity among the spin waves quickly, on timescales $O(S)$. This process is similar to what happens in phonon down-conversion processes due to interaction and it has been studied at length in the past (see \cite{alma990000176800109086}, in particular Chapter 3 by Y.~B.~Levinson).

If $\alpha\in\mathcal{L}$ and $\beta,\gamma\in\mathcal{D}$ then we have the decay of a single localized mode into two spin waves. This can happen for all $\alpha$ such that $\omega_c <\omega_\alpha < 2\omega_c$, and the spin autocorrelation function decays exponentially on timescales $O(S)$. The effective matrix element contains also some $\mathbf{Q}=\mathbf{q}_\beta+\mathbf{q}_\gamma$ dependence, and it decays as $Q\xi\gg 1$.

The last channel is when $\alpha,\beta\in \mathcal{L}$ and $\gamma\in\mathcal{D}$. In this case the matrix element is
\begin{equation}
    M_{\alpha,\beta,\gamma}\sim e^{-|\mathbf{r}_\alpha-\mathbf{r}_\beta|/\xi}/\sqrt{V},
\end{equation}
where the $\mathbf{r}$'s are the centers of localization (again dropping terms containing $q_\gamma\xi$). This channel is the only one open if the energy of the initial localized state is higher than $2\omega_c$. In this way, a localized state $\alpha$ can ``decay" into a neighboring localized state $\beta$, by emitting a spin wave of energy $\omega_\gamma=\omega_\alpha-\omega_\beta$. As long as $\omega_\gamma<\omega_c$, this process is possible and its rate can be computed using Fermi's golden rule:
\begin{equation}
    \Gamma_{\alpha\to\beta}=\pi e^{-|\mathbf{r}_\alpha-\mathbf{r}_\beta|/\xi}\frac{J^2}{S}\rho(\omega_\alpha-\omega_\beta).
\end{equation}
We can replace $\rho(\omega_\alpha-\omega_\beta)\sim \rho_0=c/(\pi J)$ with some $c=O(1)$ and obtain a rate of decay of an excitation to a neighboring excitation (therefore $e^{-|\mathbf{r}_\alpha-\mathbf{r}_\beta|/\xi}\sim 1$) with the emission of a spin wave:
\begin{equation}
    \Gamma=c\frac{J}{S}.
\end{equation}
We conclude then that {\it a single excitation} can decay to neighboring, lower energy localized states by emission of low-energy spin waves. This process can lead to two possible outcomes: either the localized state decays to another localized state with energy $\omega<2\omega_c$ and then it can decay further to localized states with energy $\omega<\omega_c$ and then eventually into spin waves; or it gets trapped because no neighbor in a radius of $\xi$ has energy $\omega<2\omega_c$. The decay out of this ``high disorder trap" can only occur through higher order processes, therefore on timescales which are $O(S^2)$ and higher.

Summarizing, the picture we obtain is the following: flipping a single spin on top of a SG ground state will create a localized excitation with a certain probability $P$ and a properly delocalized spin wave, with probability $1-P$. Some of the former localized excitations can directly decay into spin waves, while others move in the neighborhood until they either decay or get trapped between similarly high-energy localized modes and eventually will decay through higher-order processes in $1/S$ (a larger typical timescale). Since these excitations are the ones which have higher energy, this will lead to violation of the equipartition and therefore the system will be out of equilibrium. 

We plan to explore this picture quantitatively in the near future by means of more sophisticated numerical methods, which allow the treatment of interactions.

\section{Discussion}
\label{sec:conclusions}

In this work we have investigated the spin-wave excitations of the two–dimensional Heisenberg spin-glass model through a semiclassical expansion around classical equilibrium states. Our motivation was to clarify how disorder affects their dynamics and, in particular, to assess the conditions under which hydrodynamic descriptions of spin glasses may emerge despite the strong tendency toward Anderson localization in low dimensions.

At leading order in the $1/S$ expansion we obtained a quadratic spin-wave Hamiltonian whose structure reflects the correlations encoded in the underlying classical state. We showed that these correlations qualitatively modify the localization properties of the excitations with respect to those expected from uncorrelated random-matrix ensembles in the same AZ symmetry class. In the FM and AFM phases we find that the low-energy spin waves remain delocalized even in the presence of disorder, while higher-energy excitations become localized, giving rise to a mobility edge in the spectrum. In contrast, in the SG phase the correlations inherited from the classical background are RG-irrelevant and the entire spin-wave spectrum is weakly localized, consistently with the expectations for class~D systems in two dimensions.

These results have important implications for the dynamical properties of the model. In particular, localization of the non-interacting spin waves prevents a straightforward identification of the Goldstone excitations with hydrodynamic modes in the SG phase. We have argued that it is the interactions among spin waves that provide a natural mechanism to restore ergodic dynamics and a path to thermodynamics. Because the localization length of the low-energy modes grows rapidly as the energy decreases, interactions allow localized excitations to decay by emitting low energy, delocalized modes (spin waves), ultimately coupling the system to an extended low-energy sector and recovering hydrodynamic behavior at sufficiently long times.

Let us also comment on the separation of length scales that underlies our interpretation of the weak-disorder magnetically ordered regimes. As discussed above, in two dimensions an arbitrarily small density of defects may eventually destabilize FM order through the accumulation of dipolar distortions. However, the corresponding length scale $\xi_{mag}$ is expected to be exponentially large at small disorder concentration, of the form $\xi_{mag}(p) \sim e^{c/p^2}$, for some constant $c=O(1)$ \cite{Coraggio2026}. This is, if the dipoles do not end up screening themselves \cite{Parker1988}. Therefore, for weak disorder, the magnetization has an exceptionally slow flow with the system size. This should be contrasted with the behavior of the localization probes, $\phi$ and $D$, which display a much faster finite-size flow and allow one to identify localized and delocalized regions already on the system sizes accessible numerically. The mobility edge extracted in this way may therefore be a property of the pre-asymptotic, but physically relevant, magnetized regime where $ \xi_{loc}(\omega,p) \ll L \ll \xi_{mag}(p)$, where $\xi_{ loc}$ is the localization length of a spin-wave mode at frequency $\omega$. In this window of system sizes, the system behaves as magnetically ordered for all practical purposes: the magnetization is finite and changes only very slowly with $L$, while the localization properties of the spin waves are in the asymptotic, finite-size scaling regime.

The ultimate fate of the classical FM and AFM phases will be the subject of future work. On the quantum side, it would be interesting to investigate the low-energy excitations of the ground state. In this respect, numerical approaches such as DMRG could be used to analyze the excitation spectrum of the full $S=1/2$ quantum model and compare it with the semiclassical predictions, also including magnon interactions at the next orders in the $1/S$ expansion. Finally, it would be interesting to extend the present semiclassical analysis to three spatial dimensions. In $d=3$, where Anderson transitions are allowed in the AZ class~D, one may expect a genuine localization–delocalization transition already at the non-interacting level, also in the SG phase.

\section{Acknowledgments}
We thank Markus Mueller, Federico Ricci-Tersenghi, Volodya Kravtsov and Rosario Fazio for insightful discussions. G.B.T. thanks Shivaji Sondhi for the hospitality at the Rudolf Peierls Centre during the completion of this work. J.N. and G.B.T. thank Riccardo Andreoni for useful comments and collaboration on related topics. The work of A.S.\ and J.N.\ was partially funded by the European Union--NextGenerationEU under the project NRRP Project ``National Quantum Science and Technology Institute" — NQSTI, Award Number: PE00000023, Concession Decree No.~1564 of 11.10.2022 adopted by the Italian Ministry of Research, CUP J97G22000390007. This work was supported in part by the Deutsche Forschungsgemeinschaft via the cluster of excellence ctd.qmat (EXC 2147, project-id 390858490).

\clearpage
\onecolumngrid
\appendix

\section{Spin-wave Hamiltonian matrix elements}
\label{app:matrix_elements}
Let us define the tensor $\mathcal{R}_{ij}^{\alpha \beta} = R_i^{\alpha \gamma}R_j^{t,\gamma \beta}$, with $\alpha=\{x,y,z\}$. Then, the matrix elements of the spin-wave Hamiltonian in Eq.~\eqref{eq:HSW}, are given by the following expressions. For $i\neq j$:
\begin{subequations}
    \begin{align}
        &A_{ij} =  \frac{J_{ij}}{2} \left[\left( \mathcal{R}_{ij}^{xx} + \mathcal{R}_{ij}^{yy}\right) + i \left( \mathcal{R}_{ij}^{yx} - \mathcal{R}_{ij}^{xy}\right) \right] \\
        &B_{ij} = \frac{J_{ij}}{2} \left[ \left( \mathcal{R}_{ij}^{xx} - \mathcal{R}_{ij}^{yy}\right) + i \left( \mathcal{R}_{ij}^{yx} +
        \mathcal{R}_{ij}^{xy}\right) \right] \; ,
    \end{align}
\label{eq:matrixelementsOFF}
\end{subequations}
and for the diagonal elements:
\begin{equation}
A_{ii} = -\sum_{j \in \partial i}J_{ij} \mathcal{R}_{ij}^{zz} \; ,\quad
B_{ii} = 0 \; .
\label{eq:matrixelementsDIAG}
\end{equation}


\section{Dealing with zero modes}
\label{app:bogo}

\begin{figure}[b]
   \includegraphics[width=0.5\linewidth]{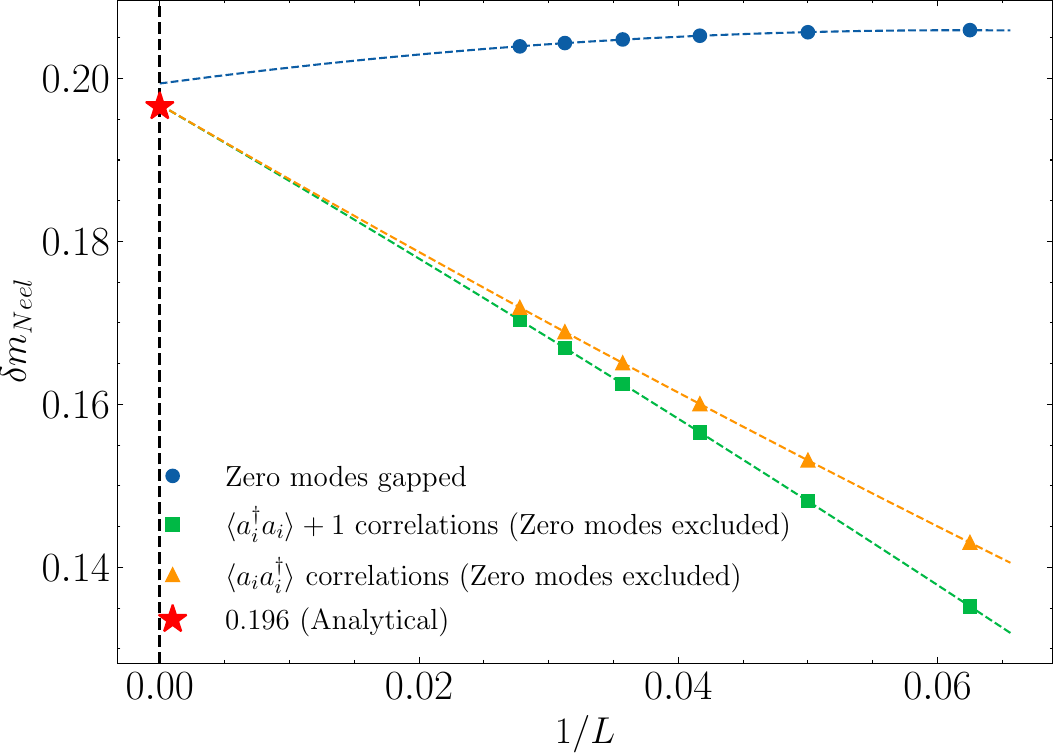}
    \caption{Leading order spin wave theory correction to the Nèel order parameter. The Nèel order parameter is $ m_{\text{Nèel}} = \frac{1}{L^2} $ and the leading order correction in the HP theory is given by $\delta m_{\text{Nèel}} =\frac{1}{S L^2} \sum_i\braket{\hat a^\dagger_i \hat a_i}$.
    The blue curve is obtained by only adding a symmetry breaking field on the first two sites of magnitude $\varepsilon = 0.5$, the yellow and green curves are obtained by manually excluding the zero modes from the sum over the quasiparticles. }
    \label{fig:Neelcorrection}
\end{figure}

The numerical implementation of the Bogoljubov transformation has to be taken with care, since it involves the diagonalization of the non-Hermitian matrix $D = \eta M$. It is possible to show that, despite being non-Hermitian, $D$  has a real spectrum, provided that the matrix $M$ is semi-positive definite \cite{blaizot_ripka_1986}, which comes as consequence of expanding around a minimum. Moreover, in the presence of disorder the only degenerate eigenspace is the kernel of the matrix, where the zero modes related to the broken symmetries live. Then, the only difficulty that one faces numerically lies in counting the number of exactly zero eigenvalues of $D$. It is worth stressing this point because, for non-normal matrices (\emph{i.e.}~$MM^\dagger \neq M^\dagger M$), the \emph{algebraic} and the \emph{geometric} multiplicity of an eigenvalue do not coincide in general. This is to be interpreted as a pathology of the eigenvector decomposition, when applied beyond the scope of its hypothesis. Indeed, by naively performing the numerical diagonalization of the dynamical matrix $D$, one finds four very small eigenvalues and would be tempted to conclude that this is the number of zero modes in the theory.

We checked numerically that the matrix $M$ has exactly three $0$ (machine precision) eigenvalues, as expected from the Goldstone mechanism. A robust check of the number of the dynamical matrix $D$ true zero modes can be done considering the singular value decomposition (SVD), whose zero singular values reveal the kernel of the matrix. We confirmed that the number of zero singular values of the dynamical matrix is exactly three, as expected. 
The extra $0$ eigenvalue of $D$ is called \emph{defective}, because its algebraic multiplicity is greater than its geometric multiplicity. It can be checked that the spurious zero eigenvector belongs to the kernel of $D^2$, that is normal and can be safely diagonalized.

Since the zero modes are associated to global rotations of the system, in principle, the Bogoljubov Hamiltonian in Eq.~\eqref{eq:Bogo-Ham generic} could be written as 
\begin{equation}
\label{eq:Bogo-zeromodes}
    H = \sum_{\alpha \in \Omega_0} \frac{\mathcal{\hat B}_\alpha \mathcal{\hat B}_\alpha}{\mathcal{I}} + \sum_{\alpha \in \Omega_+} \omega_\alpha (\hat b_\alpha^\dagger \hat b_\alpha +\mathrm{h.c.}) \; ,
\end{equation}
where $\Omega_0, \Omega_+$ are respectively the set of zero and positive modes, $\hat{\mathcal{B}}_\alpha$ is a zero mode operator and $\hat b_\alpha$ is a true quasiparticle mode. The quantity $\mathcal{I}$ is the macroscopic moment of inertia, making the contribution of the zero modes vanishingly small in the thermodynamical limit, unless they are macroscopically populated. Therefore, in the thermodynamic limit, we can treat these modes semiclassically, assigning to them a well-defined value, irrespective of their commutation relations. While this decomposition can be easily obtained in translational invariant cases, in the present case we are not able to write the Hamiltonian as in Eq.~\eqref{eq:Bogo-zeromodes} for every realization of disorder, since, as explained above, the presence of a defective eigenvalue prevents one to find all the operators $\hat{\mathcal{B}}_\alpha$. It might then seem that this must be a halting point in our analysis: these modes can not be completely excluded from the theory for a given system size $N$, since their presence is essential in order to preserve the algebra of the bosonic operator. However, we can choose the Bogoljubov vacuum in the thermodynamic limit, to have simultaneously a precise value of all 3 expectation values of $\hat{\mathcal{B}}_\alpha$, irrespective of the fact that they do not commute which allows one to exclude the contribution of these modes from the expectations values. This is equivalent to considering the zero modes as \emph{c}-numbers that can be set, for simplicity, to 0. By adopting this procedure, one is making an error on the expectation values of the observables, which vanishes in thermodynamic limit. 

The procedure can be benchmarked in the case of the antiferromagnet, where the analytical correction to the Nèel order parameter $\delta m_{\text{Nèel}} =\frac{1}{S L^2} \sum_i\braket{\hat a^\dagger_i \hat a_i}$ is known. Though at finite size $\braket{\hat a^\dagger_i \hat a_i} \neq 1 + \braket{\hat a_i \hat a^\dagger_i}$, the correct result for $\delta m_{\text{Nèel}}$ is recovered by both computing $\braket{\hat a^\dagger_i \hat a_i}$ and $1 + \braket{\hat a_i \hat a^\dagger_i}$ in the thermodynamic limit. The results of this test are shown in Fig.~\ref{fig:Neelcorrection}, where we compare this procedure to the addition of a small symmetry breaking field, which gaps the zero modes. As one can see, the two different procedures to compute the correction to the N\'{e}el order parameter tend to the same value in the thermodynamic limit, which is also reached in the case of the small external field. Since the finite size error is smaller in the previous cases than in the latter, we choose to simply exclude the zero modes from the computation of every observable. 


\section{The action of the oscillators}
\label{app:probR}

In order to compute numerically the quantity $R_\alpha(i)^2$ defined in Eq.~\eqref{eq:prob_depl}, we express it in terms of the eigenstates components $X$ and $Y$, see Eqs.~\eqref{eq:BogandInv}. Let us compute the following expectation value:
\begin{equation}
\begin{split}
    \braket{\alpha|\hat a^\dagger_i \hat a_i | \alpha} &=\sum_{\beta \gamma} \braket{\alpha|( Y^t_{i \beta} \hat b_\beta + X^\dagger_{i \beta} \hat b ^\dagger_\beta)( X^t_{i \gamma} \hat b_\gamma + Y^\dagger_{i \gamma} \hat b ^\dagger_\gamma)| \alpha } \\
    & = \sum_{\beta \gamma} Y^t_{i\beta}Y^\dagger_{i\gamma} ( \braket{0|\hat b_\alpha \hat b_\beta \hat b ^\dagger_\gamma \hat b ^\dagger_\alpha |0 } ) +  X^\dagger_{i\beta} X^t_{i\gamma} ( \braket{0|\hat b_\alpha \hat b^\dagger_\gamma \hat b_\beta \hat b ^\dagger_\alpha|0 }  ) \\
    & = |Y_{\alpha i}|^2 + |X_{ \alpha i}|^2 + \sum_{\beta} Y^t_{i \beta} Y_{i \beta} \; .
    \end{split}
\end{equation}
In the second line, we retain only the zero charge four-point functions, which are the only non-vanishing ones, and then we take the Wick contractions. Now, since $\sum_{\beta} Y^t_{i \beta} Y_{i \beta} = \braket{0 |\hat a^\dagger_i \hat a_i| 0}$ (see, e.g., the Supplemental Material of \cite{viteritti2025}), we have:
\begin{equation}
    \braket{\alpha|\hat a^\dagger_i \hat a_i | \alpha} - \braket{0 |\hat a^\dagger_i \hat a_i| 0} = |Y_{\alpha i}|^2 + |X_{ \alpha i}|^2 \; .
\end{equation}
Therefore, we identify the quantity  $R_\alpha(i)^2$ with 
\begin{equation}
    R_\alpha(i)^2 = \mathcal{R}^{-1}_\alpha \left( |Y_{\alpha i}|^2 + |X_{ \alpha i}|^2 \right) \; ,
\end{equation}
where $\mathcal{R}_\alpha$ is a proper normalization constant. Notice that $\sum_i |Y_{\alpha i}|^2 + |X_{ \alpha i}|^2$ is not the normalization of the Bogoljubov mode $\alpha$, since in the bosonic case Bogoljubov modes are normalized with respect to the metric $\eta$, see Eq.~\eqref{eq:gammaetamatr}, which gives $\sum_i |Y_{\alpha i}|^2 - |X_{ \alpha i}|^2$. 

We refer to $R_\alpha(i)^2$ as an \emph{action} increase of the harmonic oscillator on site $i$. The reason why this quantity is indeed an action becomes clear if one expresses it in terms of position and momentum variables: $\hat x_i \equiv \hat S_i^x/\sqrt{S}$ and $\hat p_i = \hat S_i^y/\sqrt{S}$. Since $\hat a_i^\dagger = \tfrac{1}{\sqrt{2}} (\hat x_i - i \hat p_i)$ and $\hat a_i = \tfrac{1}{\sqrt{2}} (\hat x_i + i \hat p_i)$, we have 
\begin{equation}
    \hat a_i^\dagger \hat a_i = \frac{\hat x_i^2 + \hat p_i^2}{2} - \frac{1}{2} \; .
\end{equation}
Hence, $R_\alpha(i)^2$ is the phase space volume projected on the state $\alpha$, reduced by the the same quantity projected on the vacuum state.

\section{Mobility edge in the FM and AFM phases}
\label{app:mobedge}

\begin{figure*}[h]
\centering
\begin{subfigure}{.45\textwidth}
  \centering
\includegraphics[width=0.98\linewidth]{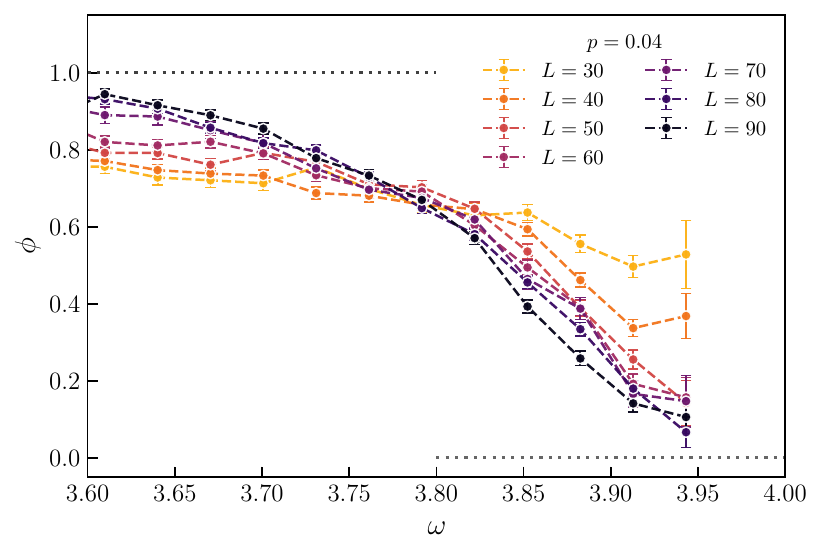}
\end{subfigure}
\begin{subfigure}{.45\textwidth}
  \centering
\includegraphics[width=0.98\textwidth]{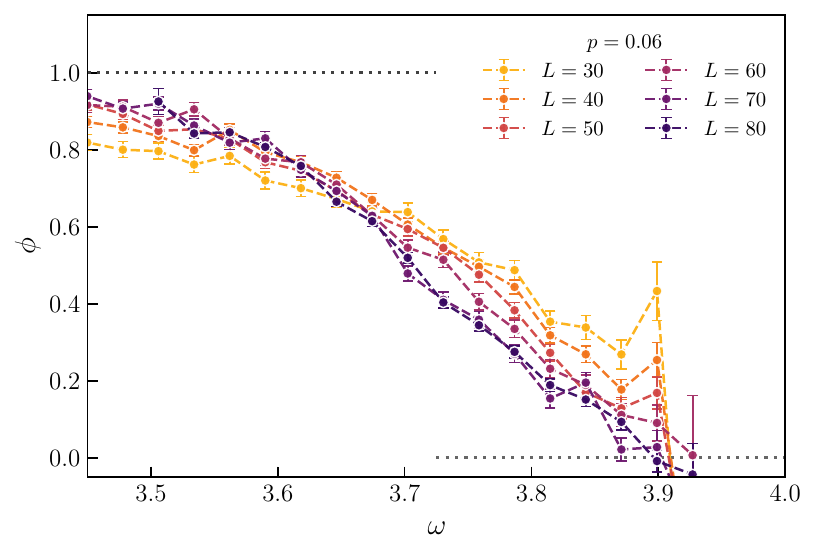}
\end{subfigure}
\begin{subfigure}{.45\textwidth}
  \centering
\includegraphics[width=0.98\linewidth]{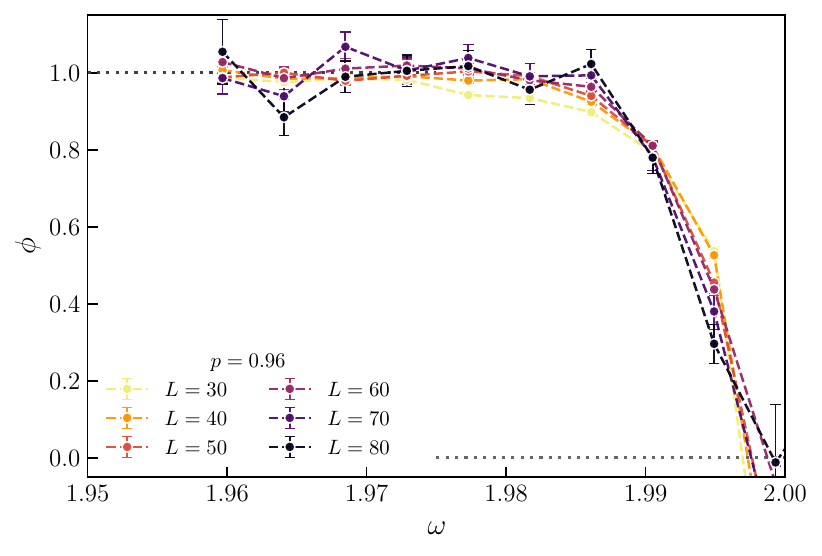}
\end{subfigure}
\begin{subfigure}{.45\textwidth}
  \centering
\includegraphics[width=0.98\textwidth]{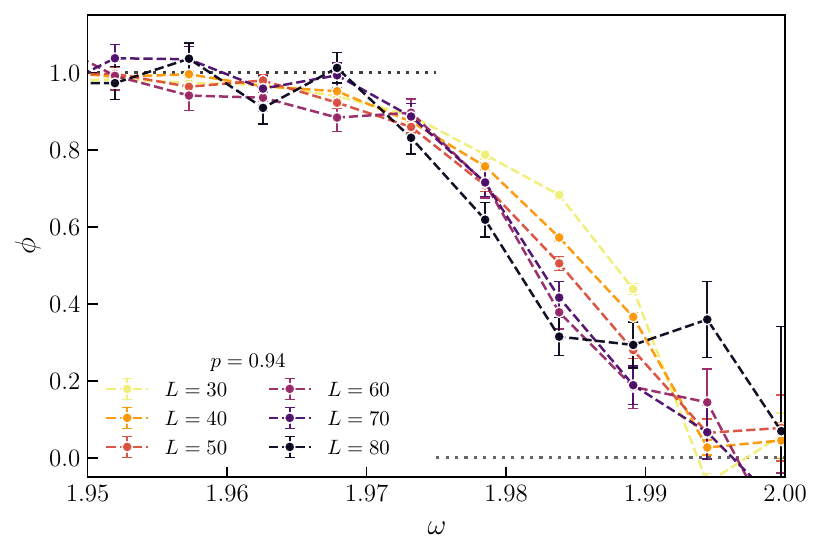}
\end{subfigure}
\caption{Rescaled $r$-parameter for values of the antiferromagnetic bond concentration $p \in \{ 0.04,0.06, 0.96,0.94\}$. For increasing value of $p$ and of $(1-p)$ the crossing shifts towards smaller energies: $\omega_c(p=0.04) \simeq 3.8$, $\omega_c(p=0.06) \simeq 3.65$, $\omega_c(p=0.96) \simeq 1.99$, $\omega_c(p=0.94) \simeq 1.97$ signaling the progressive localization of the whole spectrum moving away from the clean cases.}
\label{fig:phi_lowp}
\end{figure*}

In this section we provide further information on the mobility edge for values of $p$ closer to the clean FM/AFM states. Let us briefly comment that, in this regime, finding the classical ground state is relatively easier compared to moderate values of $p$, allowing to push the numerics to larger system sizes; however, most of the spectrum is delocalized (we are considering a small deformation of a classically ordered state), so performing a finite size scaling analysis is not always possible. In Fig.~\ref{fig:phi_lowp} we show the rescaled $r-$parameter $\phi$ for $p \in \{ 0.04,0.06,0.94,0.96\}$, computed over the largest energies of the Bogoljubov spectrum, where the crossing is expected. The phase diagram of Fig \ref{fig:phasediag} is built from the estimated values of $\omega_c(p)$. 

\begin{figure}
    \centering
    \includegraphics[width=0.6\linewidth]{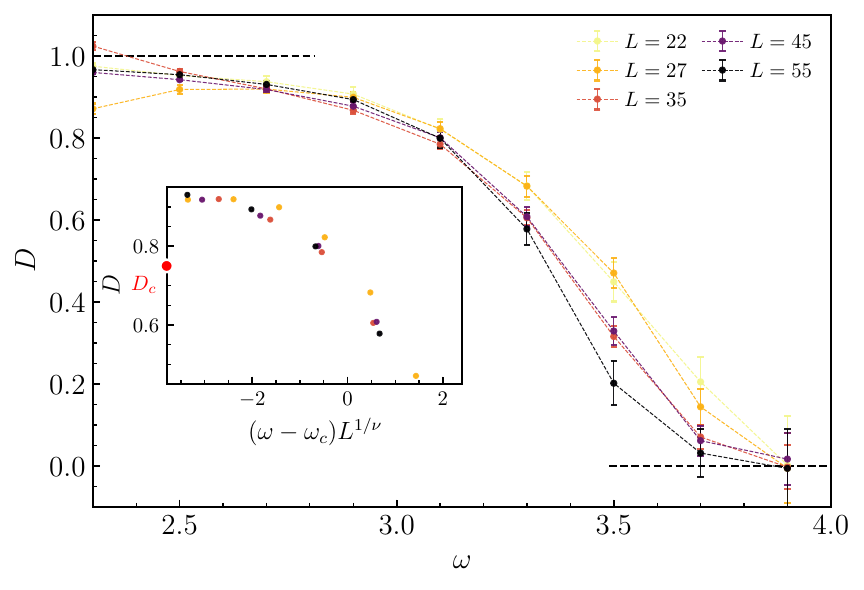}
ah     \caption{Fractal dimension $D$ for $p=0.1$, as a function of the energy $\omega$ close to the mobility edge. In the inset data are collapsed with $\omega_c = 3.2,~\nu = 2.1$, as in Fig.~\ref{fig:RGSCaling}. This fit yields a value $D_c \simeq 0.75$.}
    \label{fig:Dp01_mobilityedge}
\end{figure}

We have also computed the fractal dimension as the discrete derivative of the participation entropy for $p=0.1$. Fig.~\ref{fig:Dp01_mobilityedge} shows $D$ for energy values close to the mobility edge, while the inset presents a data collapse using the values of $(\omega_c,\nu)$ obtained from the scaling variable $\phi$ (see main text). The collapse is consistent, lending further support to our procedure. From this analysis, we estimate the fractal dimension at the critical point to be $D_c \simeq 0.75$.

\section{Finite size scaling of the fractal dimension in the SG phase}
\label{app:FractalDim}
We now describe in detail the procedure used to compute the fractal dimension $D$ and its associated $\beta$ function. The dominant source of uncertainty in the fractal dimension does not arise from the finite number of disorder realizations, but rather from sampling system sizes linearly instead of logarithmically. For this reason, it is important to consider and benchmark different methods for extracting the fractal dimension.

We first partition the participation entropy data into blocks containing $n_B$ system sizes. Within each block, we perform a linear fit of the form
\begin{equation}
S(\omega,L) = D(\omega,L(n_B)) \ln L^2 + c(\omega,L(n_B)) , ,
\end{equation}
where $L(n_B)$ denotes the geometric mean of the system sizes within the block. We have verified that the fractal dimension value extracted as the coefficient of the linear terms depends only weakly on the choice of $n_B$. Consequently, this procedure primarily smooths the data and reduces the associated uncertainty, compared to estimating $D$ via discrete derivatives of $S$.

\begin{figure}
    \centering
    \includegraphics[width=0.6\linewidth]{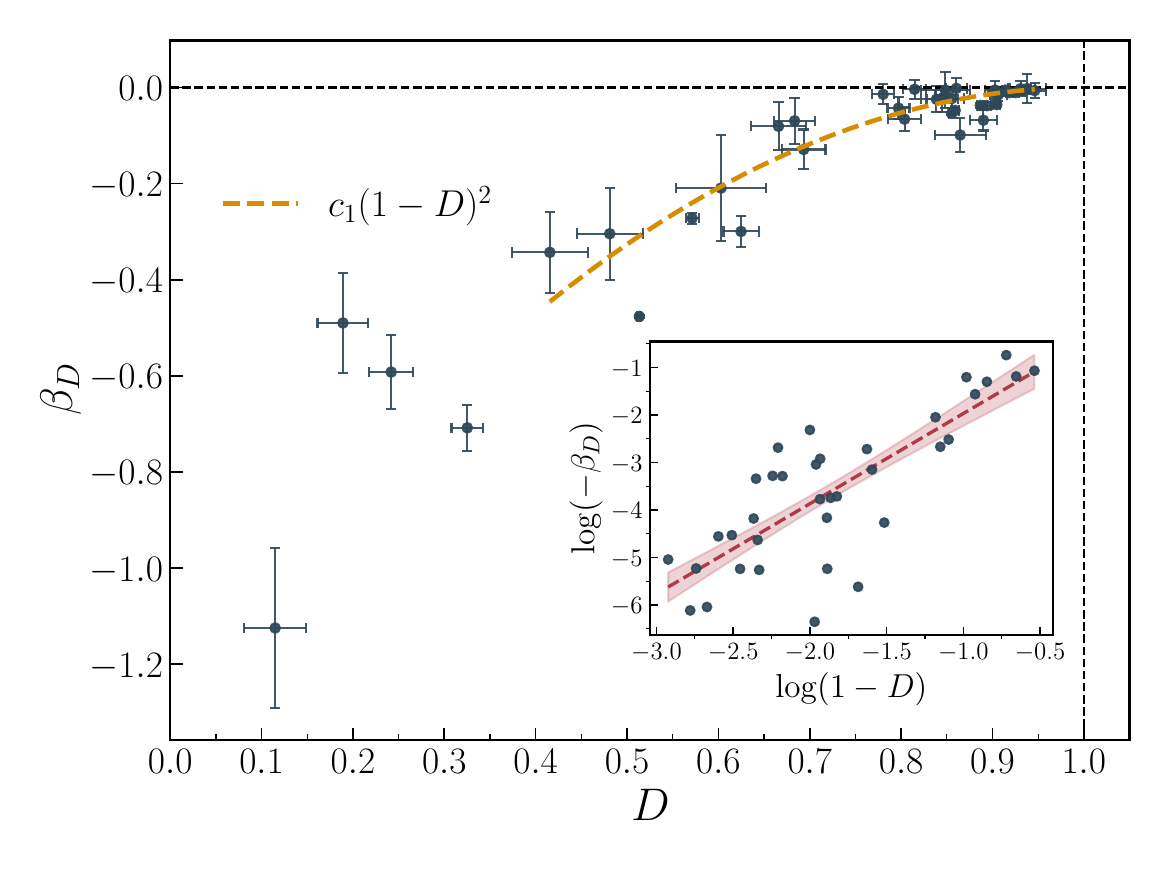}
    \caption{Parametric plot of the $\beta_D$ function as a function of the fractal dimension $D$, obtained from the quadratic fit in Eq.~\eqref{eq:Squad} with $n_B=4$. The analysis is restricted to system sizes $L \geq 22$ in order to reduce finite-size effects. The inset shows the same data on a log-log scale for small values of $1-D$, together with a fit to the form $\beta_D \simeq -c_1(1-D)^{\alpha_D}$, yielding an exponent $\alpha_D = 1.9(2)$ (with the shaded region indicating the $1\sigma$ confidence interval). The yellow dashed line in the main panel corresponds instead to the phenomenological assumption $\alpha_D = 2$, from which we extract $c_1 = 1.3(1)$.}
    \label{fig:Beta_inset}
\end{figure}

Although this approach yields cleaner data for the fractal dimension (see Fig.~\ref{fig:DandR} in the main text), it may introduce a systematic bias. In particular, a linear fit neglects the contribution of the $\beta$ function, which enters at second order and is expected to be negative. To account for this effect, we have generalized the procedure by using a quadratic fit of the form
\begin{equation}
S(\omega,L) = S_0 + D(\omega,L) \left( \ln L^2 - \ln L^2_{min} \right) + \frac{1}{2} \beta_D(\omega, L) D(\omega,L) \left( \ln L^2 - \ln L_{min}^2\right)^2 \; ,
\label{eq:Squad}
\end{equation}
where $L_{min}$ is the smallest system size in the block. While this approach leads to noisier estimates of the fractal dimension, it allows for a direct extraction of the $\beta_D$ function from the coefficient of the quadratic term. In turn, this provides a check that the scaling relation $\alpha_D = 2$ is consistent with our data. In Fig.~\ref{fig:Beta_inset}, we report the corresponding parametric plot $\beta(D)$. In the inset we extract the exponent $\alpha_D$ from a log-log fit of $\beta_D$ vs $D$, yielding a value of $\alpha_D=1.9(2)$, indeed compatible with 2.

\clearpage
\twocolumngrid
\bibliography{ref}

\begin{thebibliography}{73}%
\makeatletter
\providecommand \@ifxundefined [1]{%
 \@ifx{#1\undefined}
}%
\providecommand \@ifnum [1]{%
 \ifnum #1\expandafter \@firstoftwo
 \else \expandafter \@secondoftwo
 \fi
}%
\providecommand \@ifx [1]{%
 \ifx #1\expandafter \@firstoftwo
 \else \expandafter \@secondoftwo
 \fi
}%
\providecommand \natexlab [1]{#1}%
\providecommand \enquote  [1]{``#1''}%
\providecommand \bibnamefont  [1]{#1}%
\providecommand \bibfnamefont [1]{#1}%
\providecommand \citenamefont [1]{#1}%
\providecommand \href@noop [0]{\@secondoftwo}%
\providecommand \href [0]{\begingroup \@sanitize@url \@href}%
\providecommand \@href[1]{\@@startlink{#1}\@@href}%
\providecommand \@@href[1]{\endgroup#1\@@endlink}%
\providecommand \@sanitize@url [0]{\catcode `\\12\catcode `\$12\catcode `\&12\catcode `\#12\catcode `\^12\catcode `\_12\catcode `\%12\relax}%
\providecommand \@@startlink[1]{}%
\providecommand \@@endlink[0]{}%
\providecommand \url  [0]{\begingroup\@sanitize@url \@url }%
\providecommand \@url [1]{\endgroup\@href {#1}{\urlprefix }}%
\providecommand \urlprefix  [0]{URL }%
\providecommand \Eprint [0]{\href }%
\providecommand \doibase [0]{https://doi.org/}%
\providecommand \selectlanguage [0]{\@gobble}%
\providecommand \bibinfo  [0]{\@secondoftwo}%
\providecommand \bibfield  [0]{\@secondoftwo}%
\providecommand \translation [1]{[#1]}%
\providecommand \BibitemOpen [0]{}%
\providecommand \bibitemStop [0]{}%
\providecommand \bibitemNoStop [0]{.\EOS\space}%
\providecommand \EOS [0]{\spacefactor3000\relax}%
\providecommand \BibitemShut  [1]{\csname bibitem#1\endcsname}%
\let\auto@bib@innerbib\@empty
\bibitem [{\citenamefont {Halperin}\ and\ \citenamefont {Saslow}(1977)}]{Halperin1977}%
  \BibitemOpen
  \bibfield  {author} {\bibinfo {author} {\bibfnamefont {B.~I.}\ \bibnamefont {Halperin}}\ and\ \bibinfo {author} {\bibfnamefont {W.~M.}\ \bibnamefont {Saslow}},\ }\href {https://doi.org/10.1103/PhysRevB.16.2154} {\bibfield  {journal} {\bibinfo  {journal} {Phys. Rev. B}\ }\textbf {\bibinfo {volume} {16}},\ \bibinfo {pages} {2154} (\bibinfo {year} {1977})}\BibitemShut {NoStop}%
\bibitem [{\citenamefont {Halperin}\ and\ \citenamefont {Hohenberg}(1969)}]{Halperin_1969}%
  \BibitemOpen
  \bibfield  {author} {\bibinfo {author} {\bibfnamefont {B.~I.}\ \bibnamefont {Halperin}}\ and\ \bibinfo {author} {\bibfnamefont {P.~C.}\ \bibnamefont {Hohenberg}},\ }\href {https://doi.org/10.1103/PhysRev.188.898} {\bibfield  {journal} {\bibinfo  {journal} {Phys. Rev.}\ }\textbf {\bibinfo {volume} {188}},\ \bibinfo {pages} {898} (\bibinfo {year} {1969})}\BibitemShut {NoStop}%
\bibitem [{\citenamefont {Anderson}(1958)}]{Anderson_1958}%
  \BibitemOpen
  \bibfield  {author} {\bibinfo {author} {\bibfnamefont {P.~W.}\ \bibnamefont {Anderson}},\ }\href {https://doi.org/10.1103/PhysRev.109.1492} {\bibfield  {journal} {\bibinfo  {journal} {Phys. Rev.}\ }\textbf {\bibinfo {volume} {109}},\ \bibinfo {pages} {1492} (\bibinfo {year} {1958})}\BibitemShut {NoStop}%
\bibitem [{\citenamefont {Basko}\ \emph {et~al.}(2006)\citenamefont {Basko}, \citenamefont {Aleiner},\ and\ \citenamefont {Altshuler}}]{basko2006metal}%
  \BibitemOpen
  \bibfield  {author} {\bibinfo {author} {\bibfnamefont {D.~M.}\ \bibnamefont {Basko}}, \bibinfo {author} {\bibfnamefont {I.~L.}\ \bibnamefont {Aleiner}},\ and\ \bibinfo {author} {\bibfnamefont {B.~L.}\ \bibnamefont {Altshuler}},\ }\href@noop {} {\bibfield  {journal} {\bibinfo  {journal} {Annals of physics}\ }\textbf {\bibinfo {volume} {321}},\ \bibinfo {pages} {1126} (\bibinfo {year} {2006})}\BibitemShut {NoStop}%
\bibitem [{\citenamefont {Evers}\ and\ \citenamefont {Mirlin}(2008)}]{EversMirlin_2008}%
  \BibitemOpen
  \bibfield  {author} {\bibinfo {author} {\bibfnamefont {F.}~\bibnamefont {Evers}}\ and\ \bibinfo {author} {\bibfnamefont {A.~D.}\ \bibnamefont {Mirlin}},\ }\href {https://doi.org/10.1103/RevModPhys.80.1355} {\bibfield  {journal} {\bibinfo  {journal} {Rev. Mod. Phys.}\ }\textbf {\bibinfo {volume} {80}},\ \bibinfo {pages} {1355} (\bibinfo {year} {2008})}\BibitemShut {NoStop}%
\bibitem [{\citenamefont {Sierant}\ \emph {et~al.}(2025)\citenamefont {Sierant}, \citenamefont {Lewenstein}, \citenamefont {Scardicchio}, \citenamefont {Vidmar},\ and\ \citenamefont {Zakrzewski}}]{Sierant_2025}%
  \BibitemOpen
  \bibfield  {author} {\bibinfo {author} {\bibfnamefont {P.}~\bibnamefont {Sierant}}, \bibinfo {author} {\bibfnamefont {M.}~\bibnamefont {Lewenstein}}, \bibinfo {author} {\bibfnamefont {A.}~\bibnamefont {Scardicchio}}, \bibinfo {author} {\bibfnamefont {L.}~\bibnamefont {Vidmar}},\ and\ \bibinfo {author} {\bibfnamefont {J.}~\bibnamefont {Zakrzewski}},\ }\href {https://doi.org/10.1088/1361-6633/ad9756} {\bibfield  {journal} {\bibinfo  {journal} {Reports on Progress in Physics}\ }\textbf {\bibinfo {volume} {88}},\ \bibinfo {pages} {026502} (\bibinfo {year} {2025})}\BibitemShut {NoStop}%
\bibitem [{\citenamefont {Viteritti}\ \emph {et~al.}(2024)\citenamefont {Viteritti}, \citenamefont {Rende}, \citenamefont {Bracci-Testasecca}, \citenamefont {Niedda}, \citenamefont {Moessner}, \citenamefont {Carleo},\ and\ \citenamefont {Scardicchio}}]{viteritti2025}%
  \BibitemOpen
  \bibfield  {author} {\bibinfo {author} {\bibfnamefont {L.~L.}\ \bibnamefont {Viteritti}}, \bibinfo {author} {\bibfnamefont {R.}~\bibnamefont {Rende}}, \bibinfo {author} {\bibfnamefont {G.}~\bibnamefont {Bracci-Testasecca}}, \bibinfo {author} {\bibfnamefont {J.}~\bibnamefont {Niedda}}, \bibinfo {author} {\bibfnamefont {R.}~\bibnamefont {Moessner}}, \bibinfo {author} {\bibfnamefont {G.}~\bibnamefont {Carleo}},\ and\ \bibinfo {author} {\bibfnamefont {A.}~\bibnamefont {Scardicchio}},\ }\href {https://arxiv.org/abs/2507.05073} {\bibfield  {journal} {\bibinfo  {journal} {arXiv:2507.05073}\ } (\bibinfo {year} {2024})}\BibitemShut {NoStop}%
\bibitem [{\citenamefont {Bray}\ and\ \citenamefont {Moore}(1981)}]{Bray_1981}%
  \BibitemOpen
  \bibfield  {author} {\bibinfo {author} {\bibfnamefont {A.~J.}\ \bibnamefont {Bray}}\ and\ \bibinfo {author} {\bibfnamefont {M.~A.}\ \bibnamefont {Moore}},\ }\href {https://doi.org/10.1088/0022-3719/14/19/013} {\bibfield  {journal} {\bibinfo  {journal} {Journal of Physics C: Solid State Physics}\ }\textbf {\bibinfo {volume} {14}},\ \bibinfo {pages} {2629} (\bibinfo {year} {1981})}\BibitemShut {NoStop}%
\bibitem [{\citenamefont {Anderson}(1952)}]{Anderson_1952}%
  \BibitemOpen
  \bibfield  {author} {\bibinfo {author} {\bibfnamefont {P.~W.}\ \bibnamefont {Anderson}},\ }\href {https://doi.org/10.1103/PhysRev.86.694} {\bibfield  {journal} {\bibinfo  {journal} {Phys. Rev.}\ }\textbf {\bibinfo {volume} {86}},\ \bibinfo {pages} {694} (\bibinfo {year} {1952})}\BibitemShut {NoStop}%
\bibitem [{\citenamefont {Oguchi}(1960)}]{Oguchi_1960}%
  \BibitemOpen
  \bibfield  {author} {\bibinfo {author} {\bibfnamefont {T.}~\bibnamefont {Oguchi}},\ }\href {https://doi.org/10.1103/PhysRev.117.117} {\bibfield  {journal} {\bibinfo  {journal} {Phys. Rev.}\ }\textbf {\bibinfo {volume} {117}},\ \bibinfo {pages} {117} (\bibinfo {year} {1960})}\BibitemShut {NoStop}%
\bibitem [{\citenamefont {Castilla}\ and\ \citenamefont {Chakravarty}(1991)}]{Castilla1991}%
  \BibitemOpen
  \bibfield  {author} {\bibinfo {author} {\bibfnamefont {G.~E.}\ \bibnamefont {Castilla}}\ and\ \bibinfo {author} {\bibfnamefont {S.}~\bibnamefont {Chakravarty}},\ }\href {https://doi.org/10.1103/PhysRevB.43.13687} {\bibfield  {journal} {\bibinfo  {journal} {Phys. Rev. B}\ }\textbf {\bibinfo {volume} {43}},\ \bibinfo {pages} {13687} (\bibinfo {year} {1991})}\BibitemShut {NoStop}%
\bibitem [{\citenamefont {Altland}\ and\ \citenamefont {Zirnbauer}(1997)}]{AltlandZirnbauer1997}%
  \BibitemOpen
  \bibfield  {author} {\bibinfo {author} {\bibfnamefont {A.}~\bibnamefont {Altland}}\ and\ \bibinfo {author} {\bibfnamefont {M.~R.}\ \bibnamefont {Zirnbauer}},\ }\href {https://doi.org/10.1103/PhysRevB.55.1142} {\bibfield  {journal} {\bibinfo  {journal} {Phys. Rev. B}\ }\textbf {\bibinfo {volume} {55}},\ \bibinfo {pages} {1142} (\bibinfo {year} {1997})}\BibitemShut {NoStop}%
\bibitem [{\citenamefont {Gurarie}\ and\ \citenamefont {Chalker}(2003)}]{Gurarie2003}%
  \BibitemOpen
  \bibfield  {author} {\bibinfo {author} {\bibfnamefont {V.}~\bibnamefont {Gurarie}}\ and\ \bibinfo {author} {\bibfnamefont {J.~T.}\ \bibnamefont {Chalker}},\ }\href {https://doi.org/10.1103/PhysRevB.68.134207} {\bibfield  {journal} {\bibinfo  {journal} {Phys. Rev. B}\ }\textbf {\bibinfo {volume} {68}},\ \bibinfo {pages} {134207} (\bibinfo {year} {2003})}\BibitemShut {NoStop}%
\bibitem [{\citenamefont {Potter}\ \emph {et~al.}(2015)\citenamefont {Potter}, \citenamefont {Vasseur},\ and\ \citenamefont {Parameswaran}}]{Potter2015Universal}%
  \BibitemOpen
  \bibfield  {author} {\bibinfo {author} {\bibfnamefont {A.~C.}\ \bibnamefont {Potter}}, \bibinfo {author} {\bibfnamefont {R.}~\bibnamefont {Vasseur}},\ and\ \bibinfo {author} {\bibfnamefont {S.~A.}\ \bibnamefont {Parameswaran}},\ }\href {https://doi.org/10.1103/PhysRevX.5.031033} {\bibfield  {journal} {\bibinfo  {journal} {Phys. Rev. X}\ }\textbf {\bibinfo {volume} {5}},\ \bibinfo {pages} {031033} (\bibinfo {year} {2015})}\BibitemShut {NoStop}%
\bibitem [{\citenamefont {De~Roeck}\ and\ \citenamefont {Huveneers}(2017)}]{DeRoeck2017Stability}%
  \BibitemOpen
  \bibfield  {author} {\bibinfo {author} {\bibfnamefont {W.}~\bibnamefont {De~Roeck}}\ and\ \bibinfo {author} {\bibfnamefont {F.}~\bibnamefont {Huveneers}},\ }\href {https://doi.org/10.1103/PhysRevB.95.155129} {\bibfield  {journal} {\bibinfo  {journal} {Phys. Rev. B}\ }\textbf {\bibinfo {volume} {95}},\ \bibinfo {pages} {155129} (\bibinfo {year} {2017})}\BibitemShut {NoStop}%
\bibitem [{\citenamefont {Thiery}\ \emph {et~al.}(2018)\citenamefont {Thiery}, \citenamefont {Huveneers}, \citenamefont {M{\"u}ller},\ and\ \citenamefont {De~Roeck}}]{Thiery2018Many}%
  \BibitemOpen
  \bibfield  {author} {\bibinfo {author} {\bibfnamefont {T.}~\bibnamefont {Thiery}}, \bibinfo {author} {\bibfnamefont {F.}~\bibnamefont {Huveneers}}, \bibinfo {author} {\bibfnamefont {M.}~\bibnamefont {M{\"u}ller}},\ and\ \bibinfo {author} {\bibfnamefont {W.}~\bibnamefont {De~Roeck}},\ }\href {https://doi.org/10.1103/PhysRevLett.121.140601} {\bibfield  {journal} {\bibinfo  {journal} {Phys. Rev. Lett.}\ }\textbf {\bibinfo {volume} {121}},\ \bibinfo {pages} {140601} (\bibinfo {year} {2018})}\BibitemShut {NoStop}%
\bibitem [{\citenamefont {Morningstar}\ \emph {et~al.}(2022)\citenamefont {Morningstar}, \citenamefont {Colmenarez}, \citenamefont {Khemani}, \citenamefont {Luitz},\ and\ \citenamefont {Huse}}]{Morningstar2022Avalanches}%
  \BibitemOpen
  \bibfield  {author} {\bibinfo {author} {\bibfnamefont {A.}~\bibnamefont {Morningstar}}, \bibinfo {author} {\bibfnamefont {L.}~\bibnamefont {Colmenarez}}, \bibinfo {author} {\bibfnamefont {V.}~\bibnamefont {Khemani}}, \bibinfo {author} {\bibfnamefont {D.~J.}\ \bibnamefont {Luitz}},\ and\ \bibinfo {author} {\bibfnamefont {D.~A.}\ \bibnamefont {Huse}},\ }\href {https://doi.org/10.1103/PhysRevB.105.174205} {\bibfield  {journal} {\bibinfo  {journal} {Phys. Rev. B}\ }\textbf {\bibinfo {volume} {105}},\ \bibinfo {pages} {174205} (\bibinfo {year} {2022})}\BibitemShut {NoStop}%
\bibitem [{\citenamefont {Sandvik}(1994)}]{sandvik1994}%
  \BibitemOpen
  \bibfield  {author} {\bibinfo {author} {\bibfnamefont {A.~W.}\ \bibnamefont {Sandvik}},\ }\href {https://doi.org/10.1103/PhysRevB.50.15803} {\bibfield  {journal} {\bibinfo  {journal} {Phys. Rev. B}\ }\textbf {\bibinfo {volume} {50}},\ \bibinfo {pages} {15803} (\bibinfo {year} {1994})}\BibitemShut {NoStop}%
\bibitem [{\citenamefont {Oitmaa}\ and\ \citenamefont {Sushkov}(2001)}]{oitmaa2001}%
  \BibitemOpen
  \bibfield  {author} {\bibinfo {author} {\bibfnamefont {J.}~\bibnamefont {Oitmaa}}\ and\ \bibinfo {author} {\bibfnamefont {O.~P.}\ \bibnamefont {Sushkov}},\ }\href {https://journals.aps.org/prl/pdf/10.1103/PhysRevLett.87.167206} {\bibfield  {journal} {\bibinfo  {journal} {Physical review letters}\ }\textbf {\bibinfo {volume} {87 16}},\ \bibinfo {pages} {167206} (\bibinfo {year} {2001})}\BibitemShut {NoStop}%
\bibitem [{\citenamefont {Arrachea}\ and\ \citenamefont {Rozenberg}(2001)}]{arrachea2001}%
  \BibitemOpen
  \bibfield  {author} {\bibinfo {author} {\bibfnamefont {L.}~\bibnamefont {Arrachea}}\ and\ \bibinfo {author} {\bibfnamefont {M.~J.}\ \bibnamefont {Rozenberg}},\ }\href {https://doi.org/10.1103/PhysRevLett.86.5172} {\bibfield  {journal} {\bibinfo  {journal} {Phys. Rev. Lett.}\ }\textbf {\bibinfo {volume} {86}},\ \bibinfo {pages} {5172} (\bibinfo {year} {2001})}\BibitemShut {NoStop}%
\bibitem [{\citenamefont {Sherrington}(1977)}]{Sherrington_1977}%
  \BibitemOpen
  \bibfield  {author} {\bibinfo {author} {\bibfnamefont {D.}~\bibnamefont {Sherrington}},\ }\href {https://doi.org/10.1088/0022-3719/10/1/002} {\bibfield  {journal} {\bibinfo  {journal} {Journal of Physics C: Solid State Physics}\ }\textbf {\bibinfo {volume} {10}},\ \bibinfo {pages} {L7} (\bibinfo {year} {1977})}\BibitemShut {NoStop}%
\bibitem [{\citenamefont {Walker}\ and\ \citenamefont {Walstedt}(1977)}]{Walker1977}%
  \BibitemOpen
  \bibfield  {author} {\bibinfo {author} {\bibfnamefont {L.~R.}\ \bibnamefont {Walker}}\ and\ \bibinfo {author} {\bibfnamefont {R.~E.}\ \bibnamefont {Walstedt}},\ }\href {https://doi.org/10.1103/PhysRevLett.38.514} {\bibfield  {journal} {\bibinfo  {journal} {Phys. Rev. Lett.}\ }\textbf {\bibinfo {volume} {38}},\ \bibinfo {pages} {514} (\bibinfo {year} {1977})}\BibitemShut {NoStop}%
\bibitem [{\citenamefont {Takayama}(1978)}]{Takayama_1978}%
  \BibitemOpen
  \bibfield  {author} {\bibinfo {author} {\bibfnamefont {H.}~\bibnamefont {Takayama}},\ }\href {https://doi.org/10.1088/0305-4608/8/11/025} {\bibfield  {journal} {\bibinfo  {journal} {Journal of Physics F: Metal Physics}\ }\textbf {\bibinfo {volume} {8}},\ \bibinfo {pages} {2417} (\bibinfo {year} {1978})}\BibitemShut {NoStop}%
\bibitem [{\citenamefont {Ching}\ \emph {et~al.}(1980)\citenamefont {Ching}, \citenamefont {Huber},\ and\ \citenamefont {Leung}}]{Ching1980}%
  \BibitemOpen
  \bibfield  {author} {\bibinfo {author} {\bibfnamefont {W.~Y.}\ \bibnamefont {Ching}}, \bibinfo {author} {\bibfnamefont {D.~L.}\ \bibnamefont {Huber}},\ and\ \bibinfo {author} {\bibfnamefont {K.~M.}\ \bibnamefont {Leung}},\ }\href {https://doi.org/10.1103/PhysRevB.21.3708} {\bibfield  {journal} {\bibinfo  {journal} {Phys. Rev. B}\ }\textbf {\bibinfo {volume} {21}},\ \bibinfo {pages} {3708} (\bibinfo {year} {1980})}\BibitemShut {NoStop}%
\bibitem [{\citenamefont {Ching}\ \emph {et~al.}(1981)\citenamefont {Ching}, \citenamefont {Huber},\ and\ \citenamefont {Leung}}]{Ching_1981}%
  \BibitemOpen
  \bibfield  {author} {\bibinfo {author} {\bibfnamefont {W.~Y.}\ \bibnamefont {Ching}}, \bibinfo {author} {\bibfnamefont {D.~L.}\ \bibnamefont {Huber}},\ and\ \bibinfo {author} {\bibfnamefont {K.~M.}\ \bibnamefont {Leung}},\ }\href {https://doi.org/10.1103/PhysRevB.23.6126} {\bibfield  {journal} {\bibinfo  {journal} {Phys. Rev. B}\ }\textbf {\bibinfo {volume} {23}},\ \bibinfo {pages} {6126} (\bibinfo {year} {1981})}\BibitemShut {NoStop}%
\bibitem [{\citenamefont {Barnes}(1981)}]{Barnes_1981}%
  \BibitemOpen
  \bibfield  {author} {\bibinfo {author} {\bibfnamefont {S.~E.}\ \bibnamefont {Barnes}},\ }\href {https://doi.org/10.1103/PhysRevLett.47.1613} {\bibfield  {journal} {\bibinfo  {journal} {Phys. Rev. Lett.}\ }\textbf {\bibinfo {volume} {47}},\ \bibinfo {pages} {1613} (\bibinfo {year} {1981})}\BibitemShut {NoStop}%
\bibitem [{\citenamefont {Canisius}\ and\ \citenamefont {van Hemmen}(1981)}]{Canisius1981}%
  \BibitemOpen
  \bibfield  {author} {\bibinfo {author} {\bibfnamefont {J.}~\bibnamefont {Canisius}}\ and\ \bibinfo {author} {\bibfnamefont {J.~L.}\ \bibnamefont {van Hemmen}},\ }\href {https://doi.org/10.1103/PhysRevLett.46.1487} {\bibfield  {journal} {\bibinfo  {journal} {Phys. Rev. Lett.}\ }\textbf {\bibinfo {volume} {46}},\ \bibinfo {pages} {1487} (\bibinfo {year} {1981})}\BibitemShut {NoStop}%
\bibitem [{\citenamefont {Stinchcombe}\ and\ \citenamefont {Pimentel}(1988)}]{Stinchcombe_1988}%
  \BibitemOpen
  \bibfield  {author} {\bibinfo {author} {\bibfnamefont {R.~B.}\ \bibnamefont {Stinchcombe}}\ and\ \bibinfo {author} {\bibfnamefont {I.~R.}\ \bibnamefont {Pimentel}},\ }\href {https://doi.org/10.1103/PhysRevB.38.4980} {\bibfield  {journal} {\bibinfo  {journal} {Phys. Rev. B}\ }\textbf {\bibinfo {volume} {38}},\ \bibinfo {pages} {4980} (\bibinfo {year} {1988})}\BibitemShut {NoStop}%
\bibitem [{\citenamefont {Fava}\ \emph {et~al.}(2024)\citenamefont {Fava}, \citenamefont {Jacobsen},\ and\ \citenamefont {Nahum}}]{fava2024}%
  \BibitemOpen
  \bibfield  {author} {\bibinfo {author} {\bibfnamefont {M.}~\bibnamefont {Fava}}, \bibinfo {author} {\bibfnamefont {J.~L.}\ \bibnamefont {Jacobsen}},\ and\ \bibinfo {author} {\bibfnamefont {A.}~\bibnamefont {Nahum}},\ }\href {https://doi.org/10.1073/pnas.2401292121} {\bibfield  {journal} {\bibinfo  {journal} {Proceedings of the National Academy of Sciences}\ }\textbf {\bibinfo {volume} {121}},\ \bibinfo {pages} {e2401292121} (\bibinfo {year} {2024})},\ \Eprint {https://arxiv.org/abs/https://www.pnas.org/doi/pdf/10.1073/pnas.2401292121} {https://www.pnas.org/doi/pdf/10.1073/pnas.2401292121} \BibitemShut {NoStop}%
\bibitem [{\citenamefont {Li}\ \emph {et~al.}(2025)\citenamefont {Li}, \citenamefont {Shao},\ and\ \citenamefont {Sandvik}}]{sibei2025}%
  \BibitemOpen
  \bibfield  {author} {\bibinfo {author} {\bibfnamefont {S.}~\bibnamefont {Li}}, \bibinfo {author} {\bibfnamefont {H.}~\bibnamefont {Shao}},\ and\ \bibinfo {author} {\bibfnamefont {A.~W.}\ \bibnamefont {Sandvik}},\ }\href {https://doi.org/10.1103/PhysRevLett.134.086501} {\bibfield  {journal} {\bibinfo  {journal} {Phys. Rev. Lett.}\ }\textbf {\bibinfo {volume} {134}},\ \bibinfo {pages} {086501} (\bibinfo {year} {2025})}\BibitemShut {NoStop}%
\bibitem [{\citenamefont {Bray}\ and\ \citenamefont {Moore}(1980)}]{Bray_1980}%
  \BibitemOpen
  \bibfield  {author} {\bibinfo {author} {\bibfnamefont {A.~J.}\ \bibnamefont {Bray}}\ and\ \bibinfo {author} {\bibfnamefont {M.~A.}\ \bibnamefont {Moore}},\ }\href {https://doi.org/10.1088/0022-3719/13/24/005} {\bibfield  {journal} {\bibinfo  {journal} {Journal of Physics C: Solid State Physics}\ }\textbf {\bibinfo {volume} {13}},\ \bibinfo {pages} {L655} (\bibinfo {year} {1980})}\BibitemShut {NoStop}%
\bibitem [{\citenamefont {Kavokine}\ \emph {et~al.}(2024)\citenamefont {Kavokine}, \citenamefont {M\"uller}, \citenamefont {Georges},\ and\ \citenamefont {Parcollet}}]{Kavokine2024}%
  \BibitemOpen
  \bibfield  {author} {\bibinfo {author} {\bibfnamefont {N.}~\bibnamefont {Kavokine}}, \bibinfo {author} {\bibfnamefont {M.}~\bibnamefont {M\"uller}}, \bibinfo {author} {\bibfnamefont {A.}~\bibnamefont {Georges}},\ and\ \bibinfo {author} {\bibfnamefont {O.}~\bibnamefont {Parcollet}},\ }\href {https://doi.org/10.1103/PhysRevLett.133.016501} {\bibfield  {journal} {\bibinfo  {journal} {Phys. Rev. Lett.}\ }\textbf {\bibinfo {volume} {133}},\ \bibinfo {pages} {016501} (\bibinfo {year} {2024})}\BibitemShut {NoStop}%
\bibitem [{\citenamefont {Fernandez}(1977)}]{Fernandez_1977}%
  \BibitemOpen
  \bibfield  {author} {\bibinfo {author} {\bibfnamefont {J.~F.}\ \bibnamefont {Fernandez}},\ }\href {https://doi.org/10.1088/0022-3719/10/16/001} {\bibfield  {journal} {\bibinfo  {journal} {Journal of Physics C: Solid State Physics}\ }\textbf {\bibinfo {volume} {10}},\ \bibinfo {pages} {L441} (\bibinfo {year} {1977})}\BibitemShut {NoStop}%
\bibitem [{\citenamefont {Baity-Jesi}\ \emph {et~al.}(2015)\citenamefont {Baity-Jesi}, \citenamefont {Mart\'{\i}n-Mayor}, \citenamefont {Parisi},\ and\ \citenamefont {Perez-Gaviro}}]{Baity2015soft}%
  \BibitemOpen
  \bibfield  {author} {\bibinfo {author} {\bibfnamefont {M.}~\bibnamefont {Baity-Jesi}}, \bibinfo {author} {\bibfnamefont {V.}~\bibnamefont {Mart\'{\i}n-Mayor}}, \bibinfo {author} {\bibfnamefont {G.}~\bibnamefont {Parisi}},\ and\ \bibinfo {author} {\bibfnamefont {S.}~\bibnamefont {Perez-Gaviro}},\ }\href {https://doi.org/10.1103/PhysRevLett.115.267205} {\bibfield  {journal} {\bibinfo  {journal} {Phys. Rev. Lett.}\ }\textbf {\bibinfo {volume} {115}},\ \bibinfo {pages} {267205} (\bibinfo {year} {2015})}\BibitemShut {NoStop}%
\bibitem [{\citenamefont {Baity-Jesi}\ and\ \citenamefont {Parisi}(2015)}]{Baity2015inherent}%
  \BibitemOpen
  \bibfield  {author} {\bibinfo {author} {\bibfnamefont {M.}~\bibnamefont {Baity-Jesi}}\ and\ \bibinfo {author} {\bibfnamefont {G.}~\bibnamefont {Parisi}},\ }\href {https://doi.org/10.1103/PhysRevB.91.134203} {\bibfield  {journal} {\bibinfo  {journal} {Phys. Rev. B}\ }\textbf {\bibinfo {volume} {91}},\ \bibinfo {pages} {134203} (\bibinfo {year} {2015})}\BibitemShut {NoStop}%
\bibitem [{\citenamefont {Coraggio}\ \emph {et~al.}(2026)\citenamefont {Coraggio}, \citenamefont {Niedda}, \citenamefont {Bracci-Testasecca},\ and\ \citenamefont {Scardicchio}}]{Coraggio2026}%
  \BibitemOpen
  \bibfield  {author} {\bibinfo {author} {\bibfnamefont {A.}~\bibnamefont {Coraggio}}, \bibinfo {author} {\bibfnamefont {J.}~\bibnamefont {Niedda}}, \bibinfo {author} {\bibfnamefont {G.}~\bibnamefont {Bracci-Testasecca}},\ and\ \bibinfo {author} {\bibfnamefont {A.}~\bibnamefont {Scardicchio}},\ }\href@noop {} {\bibfield  {journal} {\bibinfo  {journal} {\emph{in preparation}}\ } (\bibinfo {year} {2026})}\BibitemShut {NoStop}%
\bibitem [{Note1()}]{Note1}%
  \BibitemOpen
  \bibinfo {note} {One can obtain an AFM minimum of the classical energy from a FM one, by performing a staggered rotation, \protect \emph {i.e.}~flipping all the spins on a sublattice, and changing sing to all the couplings, meaning that a minimum configuration at $p$ corresponds to another one at $1-p$. This symmetry is broken as soon as one takes $\hbar \protect \neq 0$.}\BibitemShut {Stop}%
\bibitem [{\citenamefont {Cieplak}\ and\ \citenamefont {Cieplak}(1985)}]{Cieplak_1985}%
  \BibitemOpen
  \bibfield  {author} {\bibinfo {author} {\bibfnamefont {M.}~\bibnamefont {Cieplak}}\ and\ \bibinfo {author} {\bibfnamefont {M.~Z.}\ \bibnamefont {Cieplak}},\ }\href {https://doi.org/10.1088/0022-3719/18/7/014} {\bibfield  {journal} {\bibinfo  {journal} {Journal of Physics C: Solid State Physics}\ }\textbf {\bibinfo {volume} {18}},\ \bibinfo {pages} {1481} (\bibinfo {year} {1985})}\BibitemShut {NoStop}%
\bibitem [{\citenamefont {Villain}(1979)}]{Villain1979}%
  \BibitemOpen
  \bibfield  {author} {\bibinfo {author} {\bibfnamefont {J.}~\bibnamefont {Villain}},\ }\href {https://doi.org/10.1007/BF01325811} {\bibfield  {journal} {\bibinfo  {journal} {Zeitschrift f{\"u}r Physik B Condensed Matter}\ }\textbf {\bibinfo {volume} {33}},\ \bibinfo {pages} {31} (\bibinfo {year} {1979})}\BibitemShut {NoStop}%
\bibitem [{\citenamefont {Binder}\ and\ \citenamefont {Young}(1986)}]{Binder_Young_1986}%
  \BibitemOpen
  \bibfield  {author} {\bibinfo {author} {\bibfnamefont {K.}~\bibnamefont {Binder}}\ and\ \bibinfo {author} {\bibfnamefont {A.~P.}\ \bibnamefont {Young}},\ }\href {https://doi.org/10.1103/RevModPhys.58.801} {\bibfield  {journal} {\bibinfo  {journal} {Rev. Mod. Phys.}\ }\textbf {\bibinfo {volume} {58}},\ \bibinfo {pages} {801} (\bibinfo {year} {1986})}\BibitemShut {NoStop}%
\bibitem [{Note2()}]{Note2}%
  \BibitemOpen
  \bibinfo {note} {In fact, \begin {equation*} {\protect \frac {1}{N^2}\DOTSB \sum@ \slimits@ _{i,j}\DOTSB \sum@ \slimits@ _{\alpha ,\beta }S^\alpha _i S^\alpha _jS^\beta _i S^\beta _j=S^4\DOTSB \sum@ \slimits@ _{\alpha ,\beta } \protect \frac {1}{3}\delta _{\alpha \beta }\protect \frac {1}{3}\delta _{\alpha \beta }=\protect \frac {S^4}{3}} \end {equation*}}\BibitemShut {NoStop}%
\bibitem [{\citenamefont {Blaizot}\ and\ \citenamefont {Ripka}(1986)}]{blaizot_ripka_1986}%
  \BibitemOpen
  \bibfield  {author} {\bibinfo {author} {\bibfnamefont {J.-P.}\ \bibnamefont {Blaizot}}\ and\ \bibinfo {author} {\bibfnamefont {G.}~\bibnamefont {Ripka}},\ }\href@noop {} {\emph {\bibinfo {title} {Quantum Theory of Finite Systems}}}\ (\bibinfo  {publisher} {MIT Press},\ \bibinfo {address} {Cambridge, Massachusetts},\ \bibinfo {year} {1986})\BibitemShut {NoStop}%
\bibitem [{\citenamefont {Nielsen}\ and\ \citenamefont {Chadha}(1976)}]{NIELSEN1976}%
  \BibitemOpen
  \bibfield  {author} {\bibinfo {author} {\bibfnamefont {H.}~\bibnamefont {Nielsen}}\ and\ \bibinfo {author} {\bibfnamefont {S.}~\bibnamefont {Chadha}},\ }\href {https://doi.org/https://doi.org/10.1016/0550-3213(76)90025-0} {\bibfield  {journal} {\bibinfo  {journal} {Nuclear Physics B}\ }\textbf {\bibinfo {volume} {105}},\ \bibinfo {pages} {445} (\bibinfo {year} {1976})}\BibitemShut {NoStop}%
\bibitem [{\citenamefont {Watanabe}\ and\ \citenamefont {Brauner}(2011)}]{Watanabe_2011}%
  \BibitemOpen
  \bibfield  {author} {\bibinfo {author} {\bibfnamefont {H.}~\bibnamefont {Watanabe}}\ and\ \bibinfo {author} {\bibfnamefont {T.~c.~v.}\ \bibnamefont {Brauner}},\ }\href {https://doi.org/10.1103/PhysRevD.84.125013} {\bibfield  {journal} {\bibinfo  {journal} {Phys. Rev. D}\ }\textbf {\bibinfo {volume} {84}},\ \bibinfo {pages} {125013} (\bibinfo {year} {2011})}\BibitemShut {NoStop}%
\bibitem [{\citenamefont {Watanabe}\ and\ \citenamefont {Murayama}(2012)}]{Watanabe_2012}%
  \BibitemOpen
  \bibfield  {author} {\bibinfo {author} {\bibfnamefont {H.}~\bibnamefont {Watanabe}}\ and\ \bibinfo {author} {\bibfnamefont {H.}~\bibnamefont {Murayama}},\ }\href {https://doi.org/10.1103/PhysRevLett.108.251602} {\bibfield  {journal} {\bibinfo  {journal} {Phys. Rev. Lett.}\ }\textbf {\bibinfo {volume} {108}},\ \bibinfo {pages} {251602} (\bibinfo {year} {2012})}\BibitemShut {NoStop}%
\bibitem [{\citenamefont {Landau}\ and\ \citenamefont {Lifshitz}(1980)}]{landlif1980}%
  \BibitemOpen
  \bibfield  {author} {\bibinfo {author} {\bibfnamefont {L.~D.}\ \bibnamefont {Landau}}\ and\ \bibinfo {author} {\bibfnamefont {E.~M.}\ \bibnamefont {Lifshitz}},\ }\href@noop {} {\emph {\bibinfo {title} {Statistical Physics, Part 1, Course of Theoretical Physics, Vol. 5}}}\ (\bibinfo  {publisher} {Pergamon Press},\ \bibinfo {year} {1980})\BibitemShut {NoStop}%
\bibitem [{\citenamefont {Bogoljubov}(1947)}]{Bogo_1947}%
  \BibitemOpen
  \bibfield  {author} {\bibinfo {author} {\bibfnamefont {N.~N.}\ \bibnamefont {Bogoljubov}},\ }\href@noop {} {\bibfield  {journal} {\bibinfo  {journal} {Journal of Physics (USSR)}\ }\textbf {\bibinfo {volume} {11}},\ \bibinfo {pages} {23} (\bibinfo {year} {1947})}\BibitemShut {NoStop}%
\bibitem [{\citenamefont {Seibold}\ \emph {et~al.}(2012)\citenamefont {Seibold}, \citenamefont {Benfatto}, \citenamefont {Castellani},\ and\ \citenamefont {Lorenzana}}]{Seibold2012}%
  \BibitemOpen
  \bibfield  {author} {\bibinfo {author} {\bibfnamefont {G.}~\bibnamefont {Seibold}}, \bibinfo {author} {\bibfnamefont {L.}~\bibnamefont {Benfatto}}, \bibinfo {author} {\bibfnamefont {C.}~\bibnamefont {Castellani}},\ and\ \bibinfo {author} {\bibfnamefont {J.}~\bibnamefont {Lorenzana}},\ }\href {https://doi.org/10.1103/PhysRevLett.108.207004} {\bibfield  {journal} {\bibinfo  {journal} {Phys. Rev. Lett.}\ }\textbf {\bibinfo {volume} {108}},\ \bibinfo {pages} {207004} (\bibinfo {year} {2012})}\BibitemShut {NoStop}%
\bibitem [{\citenamefont {Seibold}\ \emph {et~al.}(2015)\citenamefont {Seibold}, \citenamefont {Benfatto}, \citenamefont {Castellani},\ and\ \citenamefont {Lorenzana}}]{Seibold2015}%
  \BibitemOpen
  \bibfield  {author} {\bibinfo {author} {\bibfnamefont {G.}~\bibnamefont {Seibold}}, \bibinfo {author} {\bibfnamefont {L.}~\bibnamefont {Benfatto}}, \bibinfo {author} {\bibfnamefont {C.}~\bibnamefont {Castellani}},\ and\ \bibinfo {author} {\bibfnamefont {J.}~\bibnamefont {Lorenzana}},\ }\href {https://doi.org/10.1103/PhysRevB.92.064512} {\bibfield  {journal} {\bibinfo  {journal} {Phys. Rev. B}\ }\textbf {\bibinfo {volume} {92}},\ \bibinfo {pages} {064512} (\bibinfo {year} {2015})}\BibitemShut {NoStop}%
\bibitem [{\citenamefont {Xu}\ \emph {et~al.}(2020)\citenamefont {Xu}, \citenamefont {Flynn}, \citenamefont {Alase}, \citenamefont {Cobanera}, \citenamefont {Viola},\ and\ \citenamefont {Ortiz}}]{Xu2020}%
  \BibitemOpen
  \bibfield  {author} {\bibinfo {author} {\bibfnamefont {Q.-R.}\ \bibnamefont {Xu}}, \bibinfo {author} {\bibfnamefont {V.~P.}\ \bibnamefont {Flynn}}, \bibinfo {author} {\bibfnamefont {A.}~\bibnamefont {Alase}}, \bibinfo {author} {\bibfnamefont {E.}~\bibnamefont {Cobanera}}, \bibinfo {author} {\bibfnamefont {L.}~\bibnamefont {Viola}},\ and\ \bibinfo {author} {\bibfnamefont {G.}~\bibnamefont {Ortiz}},\ }\href {https://doi.org/10.1103/PhysRevB.102.125127} {\bibfield  {journal} {\bibinfo  {journal} {Phys. Rev. B}\ }\textbf {\bibinfo {volume} {102}},\ \bibinfo {pages} {125127} (\bibinfo {year} {2020})}\BibitemShut {NoStop}%
\bibitem [{\citenamefont {Mehta}(2004)}]{Mehta2004}%
  \BibitemOpen
  \bibfield  {author} {\bibinfo {author} {\bibfnamefont {M.~L.}\ \bibnamefont {Mehta}},\ }\href@noop {} {\emph {\bibinfo {title} {Random Matrices}}},\ \bibinfo {edition} {3rd}\ ed.\ (\bibinfo  {publisher} {Elsevier},\ \bibinfo {year} {2004})\BibitemShut {NoStop}%
\bibitem [{\citenamefont {Colpa}(1978)}]{Colpa_1978}%
  \BibitemOpen
  \bibfield  {author} {\bibinfo {author} {\bibfnamefont {J.}~\bibnamefont {Colpa}},\ }\href {https://doi.org/https://doi.org/10.1016/0378-4371(78)90160-7} {\bibfield  {journal} {\bibinfo  {journal} {Physica A: Statistical Mechanics and its Applications}\ }\textbf {\bibinfo {volume} {93}},\ \bibinfo {pages} {327} (\bibinfo {year} {1978})}\BibitemShut {NoStop}%
\bibitem [{Note3()}]{Note3}%
  \BibitemOpen
  \bibinfo {note} {The effective spin does not correspond to the coarse graining of a (disordered) spin configuration, but of the spins as \protect \emph {dynamical} variables, canonically conjugated to the rotation field. As such, each spin in a box of size $\lambda $ gives to the effective field an $O(1)$ contribution.}\BibitemShut {Stop}%
\bibitem [{\citenamefont {Oganesyan}\ and\ \citenamefont {Huse}(2007)}]{Oganesyan2007}%
  \BibitemOpen
  \bibfield  {author} {\bibinfo {author} {\bibfnamefont {V.}~\bibnamefont {Oganesyan}}\ and\ \bibinfo {author} {\bibfnamefont {D.~A.}\ \bibnamefont {Huse}},\ }\href@noop {} {\bibfield  {journal} {\bibinfo  {journal} {Phys. Rev. B}\ }\textbf {\bibinfo {volume} {75}},\ \bibinfo {pages} {155111} (\bibinfo {year} {2007})}\BibitemShut {NoStop}%
\bibitem [{\citenamefont {Atas}\ \emph {et~al.}(2013)\citenamefont {Atas}, \citenamefont {Bogomolny}, \citenamefont {Giraud},\ and\ \citenamefont {Roux}}]{Atas2013}%
  \BibitemOpen
  \bibfield  {author} {\bibinfo {author} {\bibfnamefont {Y.~Y.}\ \bibnamefont {Atas}}, \bibinfo {author} {\bibfnamefont {E.}~\bibnamefont {Bogomolny}}, \bibinfo {author} {\bibfnamefont {O.}~\bibnamefont {Giraud}},\ and\ \bibinfo {author} {\bibfnamefont {G.}~\bibnamefont {Roux}},\ }\href@noop {} {\bibfield  {journal} {\bibinfo  {journal} {Phys. Rev. Lett.}\ }\textbf {\bibinfo {volume} {110}},\ \bibinfo {pages} {084101} (\bibinfo {year} {2013})}\BibitemShut {NoStop}%
\bibitem [{Note4()}]{Note4}%
  \BibitemOpen
  \bibinfo {note} {From this point of view, Bogoljubov modes resemble many-body wavefunctions living in the Fock space, and this reflects the absence of particle number conservation. In fact, the same difficulties in studying eigenstate localization in real space also are found in the many-body case, even if in that case they are more severe.}\BibitemShut {Stop}%
\bibitem [{\citenamefont {Jiricek}\ \emph {et~al.}(2026)\citenamefont {Jiricek}, \citenamefont {Hopjan}, \citenamefont {Kravtsov}, \citenamefont {Altshuler},\ and\ \citenamefont {Vidmar}}]{Jiricek2026UniversalRelation}%
  \BibitemOpen
  \bibfield  {author} {\bibinfo {author} {\bibfnamefont {S.}~\bibnamefont {Jiricek}}, \bibinfo {author} {\bibfnamefont {M.}~\bibnamefont {Hopjan}}, \bibinfo {author} {\bibfnamefont {V.}~\bibnamefont {Kravtsov}}, \bibinfo {author} {\bibfnamefont {B.}~\bibnamefont {Altshuler}},\ and\ \bibinfo {author} {\bibfnamefont {L.}~\bibnamefont {Vidmar}},\ }\href {https://doi.org/10.1073/pnas.2518027123} {\bibfield  {journal} {\bibinfo  {journal} {Proceedings of the National Academy of Sciences of the United States of America}\ }\textbf {\bibinfo {volume} {123}},\ \bibinfo {pages} {e2518027123} (\bibinfo {year} {2026})}\BibitemShut {NoStop}%
\bibitem [{\citenamefont {Vanoni}\ \emph {et~al.}(2024)\citenamefont {Vanoni}, \citenamefont {Altshuler}, \citenamefont {Kravtsov},\ and\ \citenamefont {Scardicchio}}]{vanoni2023renormalization}%
  \BibitemOpen
  \bibfield  {author} {\bibinfo {author} {\bibfnamefont {C.}~\bibnamefont {Vanoni}}, \bibinfo {author} {\bibfnamefont {B.~L.}\ \bibnamefont {Altshuler}}, \bibinfo {author} {\bibfnamefont {V.~E.}\ \bibnamefont {Kravtsov}},\ and\ \bibinfo {author} {\bibfnamefont {A.}~\bibnamefont {Scardicchio}},\ }\bibfield  {journal} {\bibinfo  {journal} {Proc. Natl. Acad. Sci. (USA)}\ }\textbf {\bibinfo {volume} {121}},\ \href {https://doi.org/10.1073/pnas.2401955121} {10.1073/pnas.2401955121} (\bibinfo {year} {2024})\BibitemShut {NoStop}%
\bibitem [{Note5()}]{Note5}%
  \BibitemOpen
  \bibinfo {note} {Notice: this is just a proxy of $z$, whose correct value can be extracted from the correlation functions as a function of time. This is beyond the scope of this work and is left for following research.}\BibitemShut {Stop}%
\bibitem [{\citenamefont {Abrahams}\ \emph {et~al.}(1979)\citenamefont {Abrahams}, \citenamefont {Anderson}, \citenamefont {Licciardello},\ and\ \citenamefont {Ramakrishnan}}]{Abrahams_1979}%
  \BibitemOpen
  \bibfield  {author} {\bibinfo {author} {\bibfnamefont {E.}~\bibnamefont {Abrahams}}, \bibinfo {author} {\bibfnamefont {P.~W.}\ \bibnamefont {Anderson}}, \bibinfo {author} {\bibfnamefont {D.~C.}\ \bibnamefont {Licciardello}},\ and\ \bibinfo {author} {\bibfnamefont {T.~V.}\ \bibnamefont {Ramakrishnan}},\ }\href {https://doi.org/10.1103/PhysRevLett.42.673} {\bibfield  {journal} {\bibinfo  {journal} {Phys. Rev. Lett.}\ }\textbf {\bibinfo {volume} {42}},\ \bibinfo {pages} {673} (\bibinfo {year} {1979})}\BibitemShut {NoStop}%
\bibitem [{\citenamefont {Altshuler}\ \emph {et~al.}(2025)\citenamefont {Altshuler}, \citenamefont {Kravtsov}, \citenamefont {Scardicchio}, \citenamefont {Sierant},\ and\ \citenamefont {Vanoni}}]{altshuler2025renormalization}%
  \BibitemOpen
  \bibfield  {author} {\bibinfo {author} {\bibfnamefont {B.~L.}\ \bibnamefont {Altshuler}}, \bibinfo {author} {\bibfnamefont {V.~E.}\ \bibnamefont {Kravtsov}}, \bibinfo {author} {\bibfnamefont {A.}~\bibnamefont {Scardicchio}}, \bibinfo {author} {\bibfnamefont {P.}~\bibnamefont {Sierant}},\ and\ \bibinfo {author} {\bibfnamefont {C.}~\bibnamefont {Vanoni}},\ }\href@noop {} {\bibfield  {journal} {\bibinfo  {journal} {Proceedings of the National Academy of Sciences}\ }\textbf {\bibinfo {volume} {122}},\ \bibinfo {pages} {e2423763122} (\bibinfo {year} {2025})}\BibitemShut {NoStop}%
\bibitem [{\citenamefont {Niedda}\ \emph {et~al.}(2025)\citenamefont {Niedda}, \citenamefont {Bracci~Testasecca}, \citenamefont {Magnifico}, \citenamefont {Balducci}, \citenamefont {Vanoni},\ and\ \citenamefont {Scardicchio}}]{niedda2025}%
  \BibitemOpen
  \bibfield  {author} {\bibinfo {author} {\bibfnamefont {J.}~\bibnamefont {Niedda}}, \bibinfo {author} {\bibfnamefont {G.}~\bibnamefont {Bracci~Testasecca}}, \bibinfo {author} {\bibfnamefont {G.}~\bibnamefont {Magnifico}}, \bibinfo {author} {\bibfnamefont {F.}~\bibnamefont {Balducci}}, \bibinfo {author} {\bibfnamefont {C.}~\bibnamefont {Vanoni}},\ and\ \bibinfo {author} {\bibfnamefont {A.}~\bibnamefont {Scardicchio}},\ }\href {https://doi.org/10.1103/gcwf-jdlr} {\bibfield  {journal} {\bibinfo  {journal} {Phys. Rev. B}\ }\textbf {\bibinfo {volume} {112}},\ \bibinfo {pages} {144201} (\bibinfo {year} {2025})}\BibitemShut {NoStop}%
\bibitem [{\citenamefont {Balducci}\ \emph {et~al.}(2025)\citenamefont {Balducci}, \citenamefont {Bracci-Testasecca}, \citenamefont {Niedda}, \citenamefont {Scardicchio},\ and\ \citenamefont {Vanoni}}]{balducci2025}%
  \BibitemOpen
  \bibfield  {author} {\bibinfo {author} {\bibfnamefont {F.}~\bibnamefont {Balducci}}, \bibinfo {author} {\bibfnamefont {G.}~\bibnamefont {Bracci-Testasecca}}, \bibinfo {author} {\bibfnamefont {J.}~\bibnamefont {Niedda}}, \bibinfo {author} {\bibfnamefont {A.}~\bibnamefont {Scardicchio}},\ and\ \bibinfo {author} {\bibfnamefont {C.}~\bibnamefont {Vanoni}},\ }\href {https://doi.org/10.1103/64m5-m9ty} {\bibfield  {journal} {\bibinfo  {journal} {Phys. Rev. B}\ }\textbf {\bibinfo {volume} {111}},\ \bibinfo {pages} {214206} (\bibinfo {year} {2025})}\BibitemShut {NoStop}%
\bibitem [{Note6()}]{Note6}%
  \BibitemOpen
  \bibinfo {note} {To cut a long story short, we consider the situation settled by the theorems of \cite {de2024absence} based on the work \cite {imbrie2016many}, but see also \cite {imbrie2017local,ros2015integrals} for the connections with the perturbation theory of \cite {basko2006metal}.}\BibitemShut {Stop}%
\bibitem [{\citenamefont {John}\ \emph {et~al.}(1983)\citenamefont {John}, \citenamefont {Sompolinsky},\ and\ \citenamefont {Stephen}}]{John1983}%
  \BibitemOpen
  \bibfield  {author} {\bibinfo {author} {\bibfnamefont {S.}~\bibnamefont {John}}, \bibinfo {author} {\bibfnamefont {H.}~\bibnamefont {Sompolinsky}},\ and\ \bibinfo {author} {\bibfnamefont {M.~J.}\ \bibnamefont {Stephen}},\ }\href {https://doi.org/10.1103/PhysRevB.27.5592} {\bibfield  {journal} {\bibinfo  {journal} {Phys. Rev. B}\ }\textbf {\bibinfo {volume} {27}},\ \bibinfo {pages} {5592} (\bibinfo {year} {1983})}\BibitemShut {NoStop}%
\bibitem [{\citenamefont {Ros}\ \emph {et~al.}(2015)\citenamefont {Ros}, \citenamefont {M{\"u}ller},\ and\ \citenamefont {Scardicchio}}]{ros2015integrals}%
  \BibitemOpen
  \bibfield  {author} {\bibinfo {author} {\bibfnamefont {V.}~\bibnamefont {Ros}}, \bibinfo {author} {\bibfnamefont {M.}~\bibnamefont {M{\"u}ller}},\ and\ \bibinfo {author} {\bibfnamefont {A.}~\bibnamefont {Scardicchio}},\ }\href@noop {} {\bibfield  {journal} {\bibinfo  {journal} {Nuclear Physics B}\ }\textbf {\bibinfo {volume} {891}},\ \bibinfo {pages} {420} (\bibinfo {year} {2015})}\BibitemShut {NoStop}%
\bibitem [{\citenamefont {De~Roeck}\ \emph {et~al.}(2016)\citenamefont {De~Roeck}, \citenamefont {Huveneers}, \citenamefont {M{\"u}ller},\ and\ \citenamefont {Schiulaz}}]{de2016absence}%
  \BibitemOpen
  \bibfield  {author} {\bibinfo {author} {\bibfnamefont {W.}~\bibnamefont {De~Roeck}}, \bibinfo {author} {\bibfnamefont {F.}~\bibnamefont {Huveneers}}, \bibinfo {author} {\bibfnamefont {M.}~\bibnamefont {M{\"u}ller}},\ and\ \bibinfo {author} {\bibfnamefont {M.}~\bibnamefont {Schiulaz}},\ }\href@noop {} {\bibfield  {journal} {\bibinfo  {journal} {Physical Review B}\ }\textbf {\bibinfo {volume} {93}},\ \bibinfo {pages} {014203} (\bibinfo {year} {2016})}\BibitemShut {NoStop}%
\bibitem [{\citenamefont {Chandran}\ \emph {et~al.}(2016)\citenamefont {Chandran}, \citenamefont {Pal}, \citenamefont {Laumann},\ and\ \citenamefont {Scardicchio}}]{chandran2016many}%
  \BibitemOpen
  \bibfield  {author} {\bibinfo {author} {\bibfnamefont {A.}~\bibnamefont {Chandran}}, \bibinfo {author} {\bibfnamefont {A.}~\bibnamefont {Pal}}, \bibinfo {author} {\bibfnamefont {C.}~\bibnamefont {Laumann}},\ and\ \bibinfo {author} {\bibfnamefont {A.}~\bibnamefont {Scardicchio}},\ }\href@noop {} {\bibfield  {journal} {\bibinfo  {journal} {Physical Review B}\ }\textbf {\bibinfo {volume} {94}},\ \bibinfo {pages} {144203} (\bibinfo {year} {2016})}\BibitemShut {NoStop}%
\bibitem [{\citenamefont {Eisenmenger}\ and\ \citenamefont {Kaplianskiĭ}(1986)}]{alma990000176800109086}%
  \BibitemOpen
  \bibfield  {author} {\bibinfo {author} {\bibfnamefont {W.}~\bibnamefont {Eisenmenger}}\ and\ \bibinfo {author} {\bibfnamefont {A.~A.}\ \bibnamefont {Kaplianskiĭ}},\ }\href@noop {} {\emph {\bibinfo {title} {Nonequilibrium phonons in nonmetallic crystals.}}},\ Modern problems in condensed matter sciences. v.16.\ (\bibinfo  {publisher} {North-Holland},\ \bibinfo {address} {Amsterdam},\ \bibinfo {year} {1986})\BibitemShut {NoStop}%
\bibitem [{\citenamefont {Parker}\ and\ \citenamefont {Saslow}(1988)}]{Parker1988}%
  \BibitemOpen
  \bibfield  {author} {\bibinfo {author} {\bibfnamefont {G.~N.}\ \bibnamefont {Parker}}\ and\ \bibinfo {author} {\bibfnamefont {W.~M.}\ \bibnamefont {Saslow}},\ }\href {https://doi.org/10.1103/PhysRevB.38.11718} {\bibfield  {journal} {\bibinfo  {journal} {Phys. Rev. B}\ }\textbf {\bibinfo {volume} {38}},\ \bibinfo {pages} {11718} (\bibinfo {year} {1988})}\BibitemShut {NoStop}%
\bibitem [{\citenamefont {De~Roeck}\ \emph {et~al.}(2024)\citenamefont {De~Roeck}, \citenamefont {Giacomin}, \citenamefont {Huveneers},\ and\ \citenamefont {Prosniak}}]{de2024absence}%
  \BibitemOpen
  \bibfield  {author} {\bibinfo {author} {\bibfnamefont {W.}~\bibnamefont {De~Roeck}}, \bibinfo {author} {\bibfnamefont {L.}~\bibnamefont {Giacomin}}, \bibinfo {author} {\bibfnamefont {F.}~\bibnamefont {Huveneers}},\ and\ \bibinfo {author} {\bibfnamefont {O.}~\bibnamefont {Prosniak}},\ }\href@noop {} {\bibfield  {journal} {\bibinfo  {journal} {arXiv:2408.04338}\ } (\bibinfo {year} {2024})}\BibitemShut {NoStop}%
\bibitem [{\citenamefont {Imbrie}(2016)}]{imbrie2016many}%
  \BibitemOpen
  \bibfield  {author} {\bibinfo {author} {\bibfnamefont {J.~Z.}\ \bibnamefont {Imbrie}},\ }\href@noop {} {\bibfield  {journal} {\bibinfo  {journal} {Journal of Statistical Physics}\ }\textbf {\bibinfo {volume} {163}},\ \bibinfo {pages} {998} (\bibinfo {year} {2016})}\BibitemShut {NoStop}%
\bibitem [{\citenamefont {Imbrie}\ \emph {et~al.}(2017)\citenamefont {Imbrie}, \citenamefont {Ros},\ and\ \citenamefont {Scardicchio}}]{imbrie2017local}%
  \BibitemOpen
  \bibfield  {author} {\bibinfo {author} {\bibfnamefont {J.~Z.}\ \bibnamefont {Imbrie}}, \bibinfo {author} {\bibfnamefont {V.}~\bibnamefont {Ros}},\ and\ \bibinfo {author} {\bibfnamefont {A.}~\bibnamefont {Scardicchio}},\ }\href@noop {} {\bibfield  {journal} {\bibinfo  {journal} {Annalen der Physik}\ }\textbf {\bibinfo {volume} {529}},\ \bibinfo {pages} {1600278} (\bibinfo {year} {2017})}\BibitemShut {NoStop}%
\end{thebibliography}%

\end{document}